\documentclass[12pt, preprint]{aastex}
\usepackage{emulateapj5}

\newcommand\asca{{\it ASCA}}
\newcommand\rosat{{\it ROSAT}}

\newcommand\psc{\ifmmode{\rm\,cm^{-2}}\else{${\rm\,cm^{-2}}$}\fi}

\slugcomment{To appear in The Astrophysical Journal, March 10, 2001}

\begin{document}
\title{The Seyfert-Starburst Connection in X-Rays. II. Results and Implications}
\author{N. A. Levenson, K. A. Weaver\altaffilmark{1}, and T. M. Heckman}
\affil{Department of Physics and Astronomy, Bloomberg Center, Johns Hopkins University, Baltimore, MD 21218}
\altaffiltext{1}{Laboratory for High Energy Astrophysics, Code 662, NASA/GSFC, Greenbelt, MD 20771}
\shorttitle{The Seyfert-Starburst Connection in X-Rays}
\shortauthors{Levenson et al.}

\begin{abstract}
We present the results of X-ray imaging and spectroscopic analysis of
a sample of Seyfert 2 galaxies that contain starbursts,
based on their optical and UV characteristics.
These composite galaxies exhibit extended, soft, thermal X-ray emission,
which we attribute to their starburst components.  
Comparing their X-ray and far-infrared properties with ordinary Seyfert
and starburst galaxies, we identify the spectral characteristics 
of their various intrinsic emission sources.
The observed far-infrared emission of the composite galaxies 
may be associated almost exclusively
with star formation, rather than the active nucleus.
The ratio of the hard X-ray luminosity to the far-infrared 
and [\ion{O}{3}] $\lambda 5007$ luminosity
distinguishes most of these composite galaxies from 
``pure'' Seyfert 2 galaxies,
while their total observed hard X-ray luminosity 
distinguishes them from ``pure'' starbursts.
The hard nuclear X-ray source is generally heavily absorbed 
($N_H > 10^{23}\psc$) in the composite galaxies.  Based on
these results, we suggest that
the interstellar medium of the nuclear starburst is a
significant source of absorption.
The majority of the sample are located in groups or are interacting with 
other galaxies,
which may trigger the starburst or 
allow rapid mass infall to the central black hole, or both.      
We conclude that starbursts are energetically important in 
a significant fraction of active galaxies, 
and starbursts and active galactic nuclei
may be part of a common evolutionary sequence.

\end{abstract}

\section{Seyfert and Starburst Galaxies}
One of the major goals of astrophysics is to elucidate the physical processes
that drove the strong cosmic evolution of the active galactic nucleus (AGN) 
and galaxy populations.
The apparent ubiquity of supermassive black holes in the nuclei of present-day
galaxies and the strong correlation between the mass of the black hole
and the velocity dispersion of its ``host'' stellar spheroid \citep{Fer00,Geb00}
implies that the creation of supermassive black holes
(presumably corresponding to a luminous QSO phase) was an integral part
of the formation of ellipticals and galactic bulges.
Thus, understanding the ``starburst-AGN connection'' is a crucial component
of modern cosmogony.

Directly studying the starburst-AGN connection at high redshift is hard. The
limitations on spatial resolution and signal-to-noise make it very difficult
to use imaging or spectroscopy to disentangle processes due to the starburst
or AGN, or to try to separate cause from effect. Thus, very little is known
about the possible role of starbursts in typical high-redshift QSOs 
\citep[cf.][]{Rid00}, 
about AGNs 
in the ``Lyman Break'' population of star-forming galaxies 
\citep[e.g.,][]{Ade00}, 
about the relative energetic importance of star-formation
and AGN in high-redshift sub-mm sources 
\citep[e.g.,][]{Bar00}, 
or about the nature of the optically-faint
(presumably distant) contributors to the cosmic X-ray background that
$Chandra$ has recently
discovered \citep{Mus00,Hor00,Gia00}.

Fortunately, we have some excellent local laboratories in which the
starburst-AGN connection can be probed in considerably more detail.
The most powerful AGNs  near enough to study in such detail are the
Seyfert galaxies. The type 2 Seyferts are particularly well-suited
to such investigations because the bright glare from the central
engine has been providentially blocked by the high gas and dust column
density of an ``obscuring torus''.   Indeed, while a standard
AGN model of accretion onto supermassive black holes 
generally  describes  the 
central engines of many Seyfert galaxies successfully
(e.g., \citealt{Miy95,Tan95}, and \citealt{Nan97}), 
starbursts are also significant, often in the same
galaxies.  

To address these issues, 
we are specifically examining the X-ray properties of a sample of 
Sy 2 galaxies that definitely contain starbursts, the
Sy2/SB composite galaxies.
We present the data and detailed analysis separately 
(Levenson, Weaver, \& Heckman 2000; Paper I)\nocite{LWH00s}
and in this work concentrate on the relationship of these
results to the broader questions of the Seyfert-starburst connection.
We summarize the physical characteristics derived 
from the spatial
and spectral modelling of the sample in \S\ref{sec:models}
and present general results in \S\ref{sec:results}.
We discuss the X-ray 
properties of this sample and compare them with other samples 
in \S\ref{sec:discuss}, 
and we summarize our conclusions in \S\ref{sec:concl}. 

\begin{figure*}[thb]
\centering
\includegraphics[angle=270,width=7in]{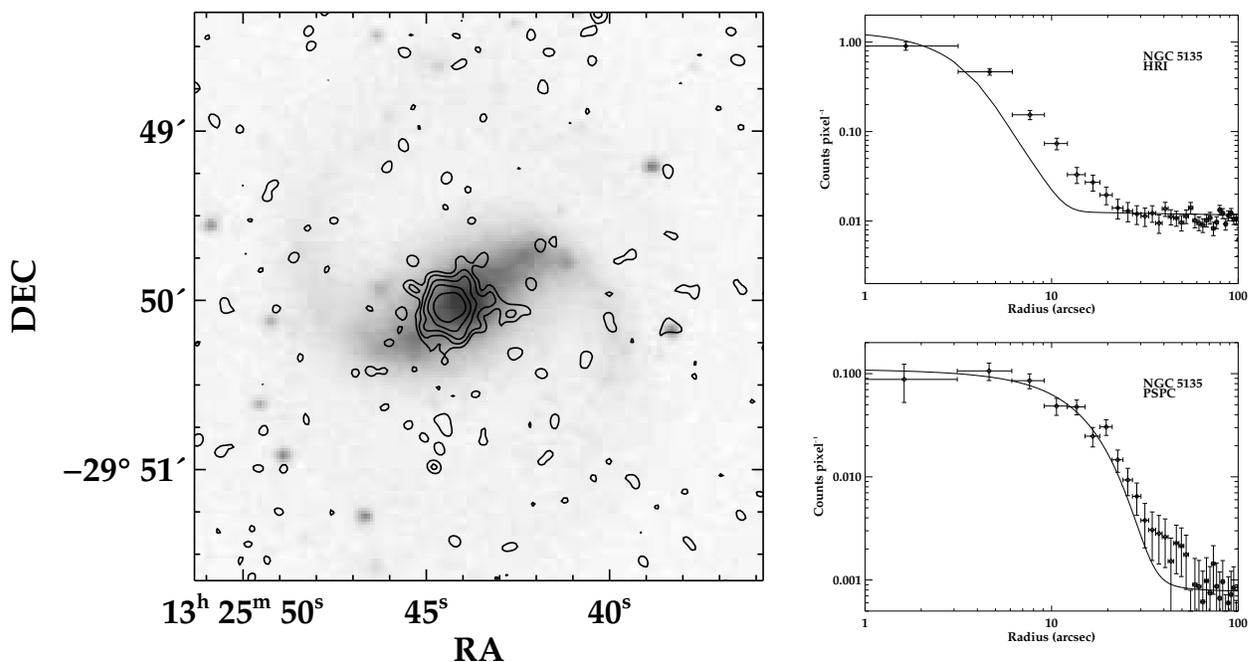}
\caption{Illustration of spatial extent of NGC 5135.
({\it left}) Contours of {\it ROSAT} HRI intensity overlaid on 
the Digitized Sky Survey optical image, where 
raw X-ray data have been smoothed by
a Gaussian of FWHM=4\arcsec.
The minimum contour level is 
3 standard deviations 
above the background, and 
contour intervals are logarithmic, in factors of 2.
({\it right}) Radial profile data points are azimuthal averages 
of counts per raw ($0\farcs5$) pixel.  The background
and amplitude are fit to the appropriate theoretical
point-spread function ({\it solid line}) for the HRI ({\it top}) and
PSPC ({\it bottom}). 
In both cases, NGC 5135 is resolved.
\label{fig:spatial}}
\end{figure*}

\section{Data and Models\label{sec:models}}
The present sample is selected from the optical and radio flux-limited
sample of \citet{Hec95}.
This larger sample contains the 30 brightest Seyfert 2 nuclei
based on [O {\sc iii}] $\lambda\lambda 4959+5007$ emission line flux
and nuclear nonthermal monochromatic flux ($\nu F_\nu$)
at 1.4 GHz from the compilation of \citet{Whi92}.
Members of the \citet{Hec95} sample have either
$\log F_{\rm [O III]} \ge -12.0 {\rm\, erg\,cm^{-2}\,s^{-1}}$,
or $\log F_{1.4} \ge -15.0 {\rm\, erg\,cm^{-2}\,s^{-1}}$, or both.     
Approximately half of this sample contain starbursts, identified
by the spectroscopic signatures of luminous young stars, such as
stellar wind lines, Balmer absorption,
and broad Wolf-Rayet emission features
\citep{Hec97,Wan97,Cid98,Gon98,Sch99,Gon00}.
We restrict our sample to these known Seyfert/starburst composites
plus several
galaxies that \citet{Hec95}
had excluded from their study only because IUE had not
observed them but which fulfill the original selection criteria.
Our sample is unbiased, but we exclude several known composite
galaxies that were not part of the original \citet{Whi92}
compilation. 
Table \ref{tab:infoir} contains 
a summary of the optical and far-infrared (FIR) properties of
our sample.  
Distances are calculated assuming
$H_0 = 75 {\rm \, km\,s^{-1}\,Mpc^{-1}}$ (column 4). Galactic column density,
{\it IRAS} fluxes in four bands, 
and galaxy inclination are noted columns 6 through 11, respectively.

As described in Paper I, we expect the composite nature of these galaxies
to be apparent in their morphologies and spectra.  While the spatial
signature of an AGN is an unresolved central point source, the
Sy2/SBs should also exhibit extended soft X-ray emission.  Analogous
to nearby starburst galaxies, the collective effect of stellar
winds and supernovae drives a hot ``superwind'' out of the galactic plane
(e.g., \citealt{Che85}, and \citealt{Hec93}),
which extends to scales of up to tens of kpc.
Spectrally, we detect X-rays due to the AGN both 
directly---though obscured---and indirectly, in scattered light.
Both of these continuum components are
characterized by a power law of photon index $\Gamma \approx 1.9$,
as observed in Seyfert 1 galaxies \citep{Nan94}, assuming the
central engines of Sy 1s and Sy 2s are intrinsically identical.  
With the addition
of a starburst, soft thermal emission ($kT \sim 1$ keV) is also present.

With this physical motivation, we model  
X-ray images and spectra obtained with \rosat\ and \asca.
The \rosat\ High Resolution Imager (HRI) provides spatial
resolution of about $5\arcsec$.  The \rosat\ Position Sensitive
Proportional Counter (PSPC) provides spatial resolution of about
$30\arcsec$, with higher sensitivity to measure the extremely
low surface brightness extended emission better, and some
spectral resolution.  Data obtained with \asca\ allow spectroscopy
at energies ranging from about 0.5 to 10 keV, which we combine with
PSPC spectra when available.

Tables \ref{tab:extend} and \ref{tab:best1} summarize the results
of the model fitting, which we describe in detail in Paper I.
We measure the fraction of resolved emission in the HRI and PSPC
images two ways.  In the first method (Table \ref{tab:extend}, columns
3 and 6), it is the residual
in excess of a single point source, which we model as the instrumental point
spread function.  This is a  conservative estimate, for it usually implies 
an unfilled ring of extended emission, which is physically unlikely.
In the more realistic method (columns 4 and 7), we constrain the
residual profile to be flat within the central core of the point
spread function. Eleven of the twelve galaxies observed with \rosat\
exhibit extended emission, measured by at least one of these methods.
As an example, Figure \ref{fig:spatial} illustrates the spatial extent 
of NGC 5135
with a \rosat\ HRI map and radial profiles from both HRI and PSPC 
observations, which represent the conservative measurement method.

Table \ref{tab:best1} contains the parameters of the best-fitting
spectral models.  In addition to the power laws and thermal
emission described above, we also considered including Fe K$\alpha$ lines
near 6.4 keV and provide upper limits on their equivalent widths
where the lines are not required.  In all cases, the inclusion of additional
model components or free parameters is significant at the 90\%
confidence level, based on an $F$ test.  With one exception (Mrk 477),
all the twelve galaxies for which we have \asca\ spectra require
a thermal component, with typical $kT = 0.8$ keV, which is characteristic
of pure starburst galaxies.  For comparison with the imaging results, we 
list the
fraction of soft (0.5--2.0 keV) X-rays that are due to this
thermal emission in column 8 of Table \ref{tab:extend}.

\section{Results\label{sec:results}}
\subsection{Synopsis of Paper I}
Based on spectral fits, we find that Seyfert/starburst composite galaxies 
possess two types of continuum shapes (Figure \ref{fig:modspec}).  
The first shows a steadily 
decreasing spectral intensity from low energies to high energies and no 
clear evidence for a hard X-ray ``bump'' that would suggest we are seeing 
a buried AGN directly through large amounts of obscuring matter (e.g., 
NGC 7130). 
This spectral type is common to nearby starburst galaxies 
\citep*{Dah98}, 
although in the Sy2/SBs the AGNs also contribute to the hard X-ray flux.  
The second shows two distinct spectral
components: a soft X-ray peak and a hard X-ray 
peak, consistent with an absorption hump associated with 
a buried AGN (e.g., Mrk 477).
This type is  
typical of pure, scattering-dominated Seyfert 2 
galaxies such as Mrk 348 \citep*{Net98,Awa00}.
The observed distribution in spectral types can be produced 
by varying contributions of an AGN and a starburst, including some 
examples that
contain both the direct and scattered AGN components in addition
to the soft X-ray starburst (e.g., Mrk 273). 
Two members of our sample are exceptional, and these representative
spectra do not characterize them well.
The complex spectrum of NGC 1068 includes two thermal and three Fe line
components in addition to the power law of the scattered AGN.  The
central source of NGC 6221 is only slightly absorbed.  Although a thermal
component is present and of comparable luminosity to the other examples,
it is weak with respect to the nuclear source. 

All the composite galaxies we have observed with \asca\ 
except NGC 6221 exhibit a
heavily absorbed ($N_H > 10^{23} {\rm\,cm^{-2}}$) power-law component,
either viewed directly or inferred from other means.
In about half the cases, we explicitly model the buried AGN and its
scattered soft X-ray emission.  In the other examples, the
models consist of thermal emission and a single power-law component.
We interpret the 
low foreground column density  
and lack of a second, absorbed (in the \asca\ bandpass) 
power-law component in NGC 1068, Mrk 1066, Mrk 266,  
and NGC 7130 as evidence that the intrinsic AGN
is completely buried and
fully absorbed by $N_H \gtrsim 10^{24} {\rm\,cm^{-2}}$.  
\asca\ observations
are not sensitive to
such high column densities, and
higher-energy data are required to measure the intrinsic power law
and absorption accurately in these instances, 
as well as in NGC 5135.
NGC 1068 has been observed with {\it BeppoSAX}, which is
sensitive up to 100 keV, and the broad-band spectrum
is consistent with the complete blocking of the intrinsic AGN \citep{Mat97}.
Furthermore,  NGC 7674, one of two members
of our sample that was not observed with \asca, is observed 
with {\it BeppoSAX} to be Compton thick \citep{Mal98}.

\begin{center}
\includegraphics[width=3.1in]{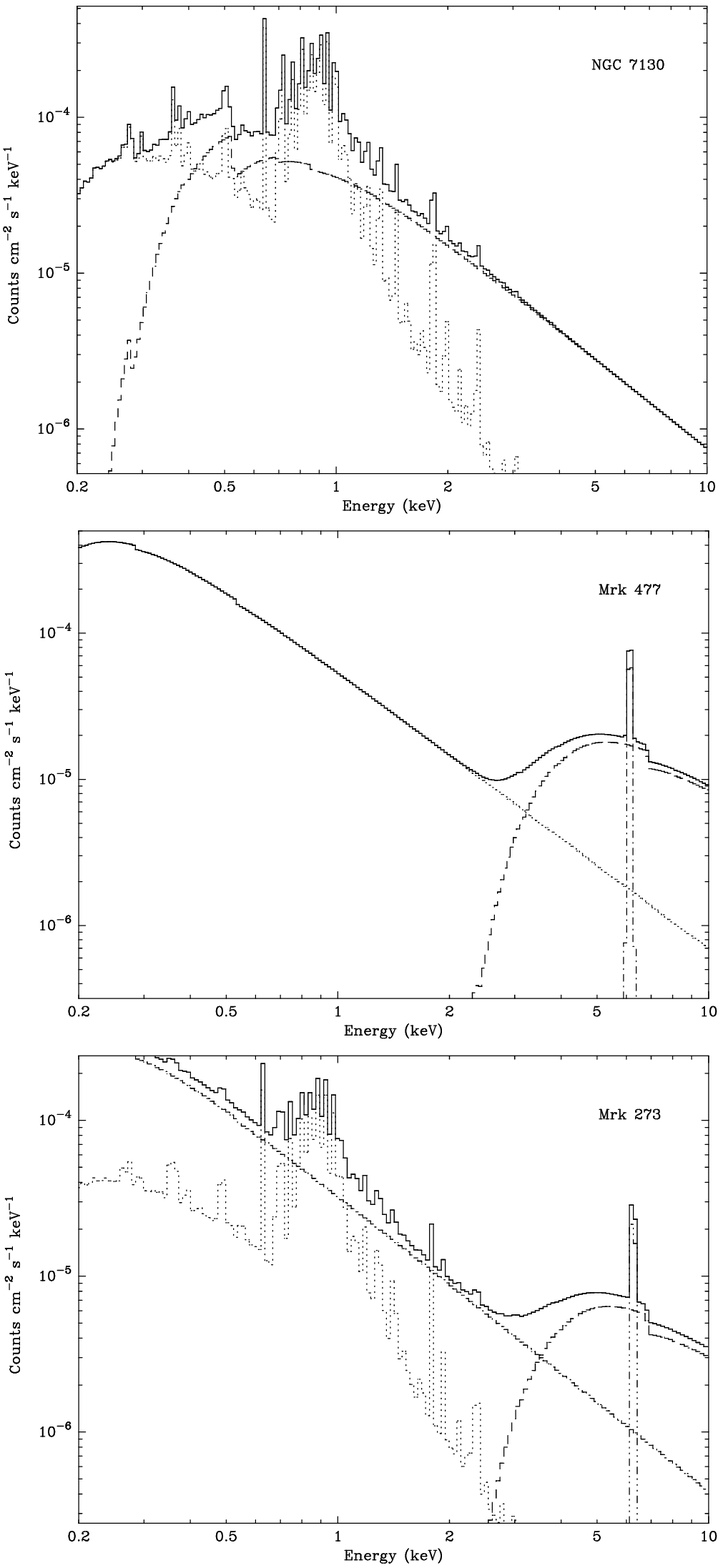}
\end{center}
\figcaption
{Best-fitting model spectra, explicitly revealing the multiple components that
comprise them.  NGC 7130 and Mrk 477 illustrate the two characteristic 
continuum shapes of the composite galaxies, depending on whether the
intrinsic AGN is completely absorbed, as in the former, or partially
emerges to add a high-energy bump, as in the latter.  Varying degrees
of absorption, scattering, and relative strength of the thermal
component (e.g., Mrk 273), account for most variations of the observed spectra.
\label{fig:modspec}
}

\subsection{X-ray Properties of the Sample}
Where both the intrinsic and scattered AGN are observed,
we estimate how much of the AGN spectrum is scattered into our
line of sight by defining a scattering fraction, $f_{scatt}$, calculated from 
the ratio of the soft X-ray power law to the hard X-ray (absorbed)
power law. 
The observed $f_{scatt}$ varies from 
0.02 to 0.11, with an average of 0.05. 
The scattered fractions are similar to ordinary 
Seyfert 2s \citep{Mul93} and polarized broad line Sy 2s \citep{Awa00}.
The scattered fraction does not depend on the column density
that absorbs the hard emission; both high and low column densities
produce high and low scattered fractions.
While broad optical lines have been observed in the polarized light
of some sample members, the measured scattering fraction in these
instances is moderate, not extreme.

With the exception of Mrk 477, all the preferred models include thermal 
emission.
The temperatures of these thermal components 
range from 0.16 to 2.3 keV, having a mean $kT= 0.82$ keV.
The temperature distribution of the thermal component of the
composite galaxies is not significantly different from the moderate
temperatures measured in starburst galaxies, as Figure \ref{fig:kthist}
illustrates, comparing with starburst spectra from \rosat\ and
\asca\ that have been analyzed jointly in the literature. 
\begin{center}
\includegraphics[width=3.5in]{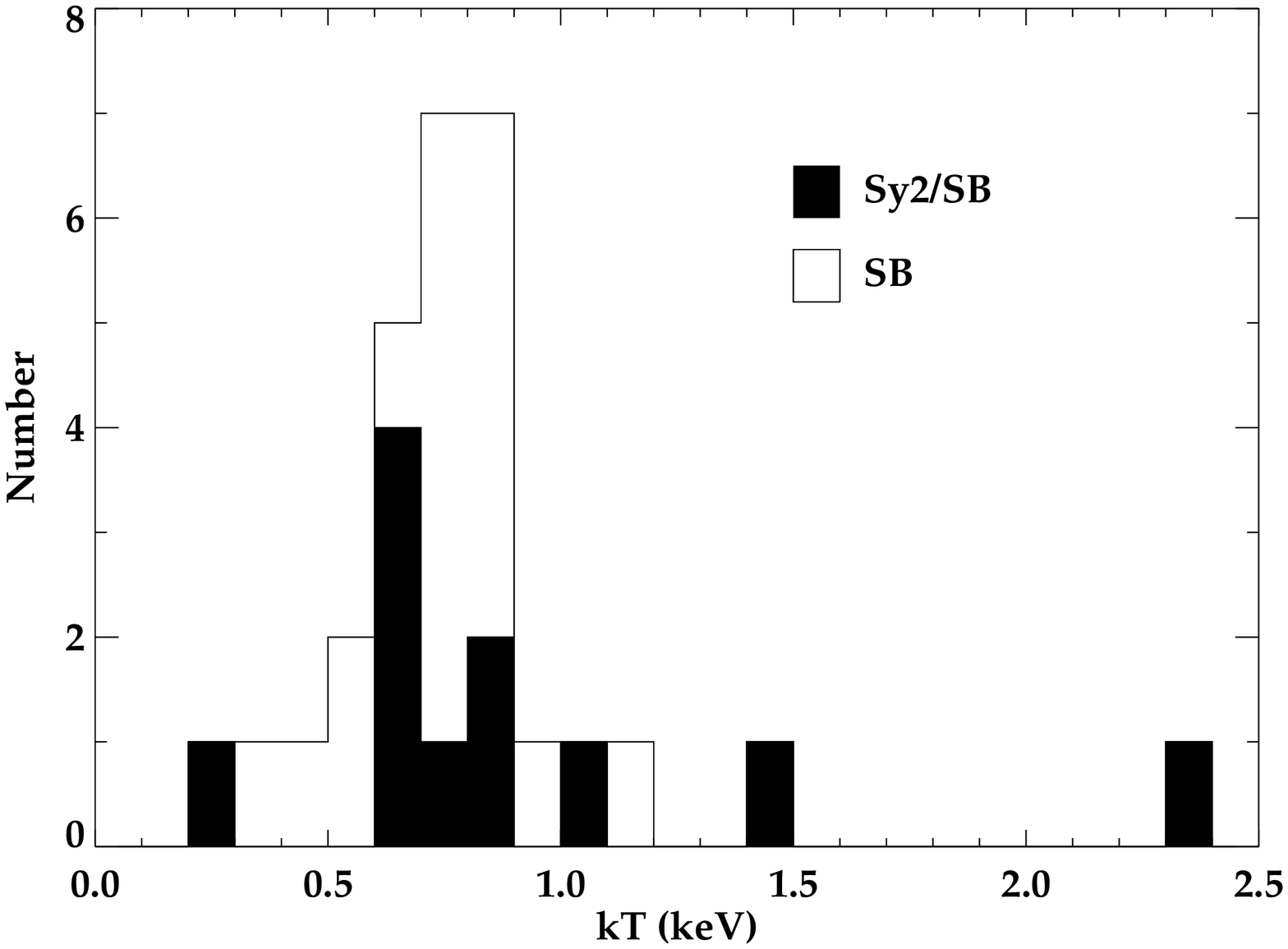}
\end{center}
\vskip -0.2in
\figcaption
{Histogram of thermal component temperatures measured in the Sy2/SB
sample and the moderate thermal temperatures of ordinary starburst
galaxies.  The mean temperatures are not significantly different
between these two samples, supporting the starburst origin
of the thermal component in the composite galaxies.
Joint \rosat+\asca\ spectra were analyzed for the starburst sample,
which include: 
NGC 253, NGC 3029, NGC 3628, NGC 4631, and M82 (Dahlem et al. 1998); 
NGC 1097, NGC 4736, and M83 (D. K. Strickland 2000, private communication);
NGC 1569 \citep{del96};
NGC 1808 \citep{Awa96}; 
NGC 2146 \citep{del99}; 
NGC 3256 \citep{Mor99}; 
NGC 3310 \citep{Zez98}; 
NGC 4038/4039 \citep{San96}; 
NGC 4449 \citep{del97}; 
NGC 6240 \citep{Iwa99};
and Arp 299 \citep{Hec99}.
\label{fig:kthist}}
\medskip

The soft X-rays in Mrk 477 
are consistent with purely scattered flux from an AGN \citep{Awa00}, but
the spectrum of Mrk 477 can also be fitted with a thermal component 
added to the best-fitting scattered power law plus line model.
The additional component is not statistically significant and the 
errors are large, but the best thermal model parameters are reasonable: $kT =0.9$ keV 
and $A3 = 9.3\times10^{-6}$. 

The fluxes listed in Table \ref{tab:best1} are the
observed quantities.  To compare the galaxies, we calculate the 
corresponding luminosities (Table \ref{tab:xlum}), corrected for
absorption by the intervening Galactic column density. 
Most of the composite galaxies have large amounts of intrinsic 
absorption, and these absorption corrections can be very uncertain.
Therefore, we consider only the emerging hard X-ray 
luminosity from 2 to 10 keV, $L_{2.0-10}$ or $HX$, not corrected
for the intrinsic column density.
We find values of $L_{2.0-10}$ between $9.1\times10^{40}$
and $3.3\times10^{42} {\rm\, erg\,s^{-1}}$,
with a mean $<L_{2.0-10}> = 4.2\times10^{41}{\rm\, erg\,s^{-1}}$.
The total soft X-ray luminosities 
between 0.5 and 2.0 keV, $L_{0.5-2.0}$ or $SX$, 
are corrected for absorption due to our Galaxy.
They range from $2.6\times10^{40}$ to 
$5.6\times10^{41}{\rm\, erg\,s^{-1}}$, with 
$<L_{0.5-2.0}> = 1.3\times10^{41}{\rm\, erg\,s^{-1}}$.
In all galaxies except Mrk 477, 
the total hard and soft luminosities are similar.
For comparison, we list the spatially extended and 
total soft X-ray flux measured with the
\rosat\ HRI and PSPC.  
The HRI provides no spectral information, and the resolved components
alone 
of the PSPC observations have inadequate signal with which to measure the
spectrum directly.  Thus, 
we convert the observed count rates to fluxes adopting an
emission model of a thermal plasma at 1 keV with Galactic absorption.  
The extended fractions are based
on the conservative estimates of resolved emission.  In Table \ref{tab:xlum}, we
also list the medium (0.5-4.5 keV) X-ray luminosities, $MX$, which we discuss below.

From our spectral decomposition models of power law and thermal emission,
we can compare the relative strength of the thermal and scattered 
components. 
$SX_{thermal}$ ranges from 0.2 to 1.2 times $SX_{scatt}$, with an average of 0.6. 
However, the data are not sensitive enough to search for a correlation.
This is partly a selection effect since a weak starburst 
would not be detected, as in Mrk 477.
The fraction of total soft X-ray luminosity due to the thermal component of
the best-fitting spectral models is listed in column 8 of 
Table \ref{tab:extend}, along with the fraction of extended emission
measured in \rosat\ images.  

The spectral and imaging results are generally consistent.
The spectral measurement of 100\% thermal soft X-ray luminosity in
IC 3639 is an overestimate, considering the evidence for some
scattered nuclear emission, although it is not statistically significant
in the current data.
Both imaging methods begin with the assumption that a single point source
is an appropriate model for the principal spatial distribution of X-rays.  
In the
examples where the X-ray emission is extremely extended or
exhibits a great deal of structure, such as NGC 1068, Mrk 266, and NGC 7582, this
assumption is incorrect, and the resulting imaging and spectral fits disagree somewhat.

In several cases with high signal-to-noise ratios, the differences 
between the softest (0.2--0.4 keV) and harder (0.4-2.0 keV) emission
that the PSPC spectrally resolves also supports the 
identification of extended thermal plasma.  Emission in these two
broad bands is spatially distinct (Paper I, Figures 5--8).  The
harder emission tends to be more centrally concentrated, and most of it
is due to the AGN directly.  The softer flux is more diffuse.  
Although some of this emission comes from the AGN, more of it is due to
the extended component, which we associate with the starburst-driven outflow.

The greatest disagreement between the imaging and spectroscopic identification
of the extended  thermal component arises in the obviously variable 
members of the sample. The \rosat\ and \asca\ observations of each of these
galaxies, NGC 6221 and NGC 7582, were separated by a year or two.
We account for these discrepancies by assuming the intrinsic AGN or its
obscuring column density changed and do not require any variation of the extended
thermal component itself. 

We adopt solar abundances in the spectral models because the data are
inadequate to reliably measure the metallicity. 
Single-temperature thermal plasma models of 
high-quality spectra of starburst galaxies often require extremely low
abundances, however \citep{Dah98}.  In these models, with a reduced population of metal ions,
the X-ray line emission relative to the continuum is suppressed.  The success of the model
fits does not necessarily mean that the abundance is genuinely low in the X-ray-emitting
region, but rather that the observed line emission is diluted with respect to the continuum,
possibly due to the presence of an additional continuum source 
\citep{Wea00c,Str00}.
In our models of Sy2/SBs, the thermal component represents
the net starburst emission.  In light of the empirical conclusions in 
pure starburst galaxies, we examined fixed low abundance ($Z=0.05Z_\odot$)
thermal models in the best-fitting cases.  Table \ref{tab:lowz} summarizes
these results in terms of two key parameters: the ratio of intrinsic to scattered
AGN continuum, and the fraction of soft emission that is thermal.  
We apply the low abundance to the best-fitting model and note the quality
of this fit ($\chi^2/$dof) in the last column.
In general, 
the low abundance models do not significantly alter the observed scattered
fractions, but they tend to increase the proportion of thermal emission.
The effects are strongest in Mrk 273, Mrk 463, and NGC 6221, in which the thermal
fraction increases significantly and the scattered fraction also 
decreases measurably.  In Mrk 273, in fact, the scattered component 
is not directly detected in the low abundance model.

\section{Discussion\label{sec:discuss}}
\subsection{Composite X-ray Sources\label{subsec:xray}}

All the sample galaxies exhibit X-ray evidence 
for an AGN, which is consistent with their optical classification as 
Seyfert 2s.  NGC 6221 is unique in that it is the
only galaxy that has a truly Seyfert 1-like X-ray spectrum, with a 
broadened Fe K$\alpha$ line that suggests origin in an accretion disk
\citep{Nan97}.  The other objects contain a heavily
absorbed hard X-ray component consistent with a buried AGN 
($N_H > 10^{23}\psc$), similar 
to more traditional X-ray selected Seyfert 2s.   
Where the intrinsic hard X-ray emission is completely absorbed and the
AGN is not directly detected, the  Fe K$\alpha$ line emission is consistent 
with scattered light from an AGN (see Paper I, Figure 12).   
At high energies, the Sy2/SB composites
are indistinguishable from scattering-dominated Seyfert 2s 
\citep{Net98,Awa00}.  The soft X-ray images and spectra, however, are
complex.  The data require composite 
energy sources---both an AGN and a starburst---in each case.

\subsection{X-ray Morphology}
The PSPC and HRI images show extended features. Eleven of twelve observed galaxies 
are extended in the HRI (according to at least one of two measurement techniques) 
while six of eight are significantly extended in the PSPC.  Extended emission is visible 
to scales as large as about 30 kpc, 
and the fraction of flux in the extent can 
make up as much as 80\% of the total flux.  The presence of extended
emission on such large scales 
rules out the central AGN as the exclusive source of soft X-rays. 
The most likely sources of soft X-rays are the galaxy disk, AGN-driven jets
or winds, and starburst-driven winds.  We consider these possibilities in
turn, ultimately favoring the starburst origin of extended emission in
these composite galaxies.

The fraction of soft X-ray emission that is resolved in images
is comparable to the fraction of thermal emission we detect
spectroscopically.  Thus, we attribute a common origin to
the extended soft X-rays and the thermal component.
Although the interstellar medium (ISM) of galaxies and AGN outflows share
some of these characteristics, overall these 
extended/thermal components most resemble starburst galaxies
in their physical properties, including sizes, temperatures, and
X-ray to far-infrared luminosity ratios, as we discuss below.

Normal galaxies emit X-rays.  These originate in point 
sources---including individual stars, X-ray binaries and 
supernovae---and a hot diffuse component of the ISM, 
which is the net result of supernova remnants.
The total X-ray luminosity of a normal galaxy is
generally less than $10^{40} {\rm\,erg\, s^{-1}}$ 
\citep[e.g.,][]{Fab96}, 
which is
significantly smaller than the extended emission we measure
in the Sy2/SB sample.  Also, for each composite galaxy,
the optical and X-ray morphology are distinct.  Particularly in the
nearby examples, where instrumental resolution is not a significant
limitation, we find that the X-ray emission poorly traces the
optical disk.  Instead, in the nearly face-on cases,
it is strongly concentrated near the nucleus,
although it is resolved and therefore not due entirely to the AGN.  
In the edge-on cases, the soft X-rays extend perpendicular to the
plane and are not confined to the disk.
Thus,
normal galactic emission is not the source of the diffuse X-rays
in the Sy2/SB sample.

With an outflow or ``jet,''  an AGN can directly generate 
extended, thermal X-ray emission, shock-heating the ambient material.
(We disregard direct generation of non-thermal X-rays in the jet, which would not share
the thermal spectral characteristic we observe.)
AGN-driven jets exist in nearby radio-quiet AGN on scales as 
large as a few kpc \citep{Col96}.
In Seyfert galaxies, however, AGN-driven outflows tend to be
small, having physical scales less than 500 pc \citep{Ulv84} 
rather than extending to the tens of kpc we observe in these examples.
In a small sample of Seyfert galaxies, \citet{Bau93}
observe kpc-scale radio extent and argue that this is due to
circumnuclear starbursts, based on the alignment of this emission
with the galaxies' minor axes and its misalignment with smaller-scale
radio emission.  

A  radio jet of 500 pc radius is observed in NGC 1068 
\citep{Wil82}.
This linear radio structure is much smaller than the X-ray extent 
this galaxy, however, so the extended and thermal X-ray emission
of NGC 1068 are not due to the jet.
At radio wavelengths, Mrk 78 
extends around a central core, linearly toward the east
and in a diffuse lobe toward the west, 
with associated [\ion{O}{3}] emission
\citep{Ped89}. The radio size is around 2\arcsec\ (1.4 kpc),
which is much smaller than the X-ray radius of 20--30 kpc.
Another distinction is that while the radio and optical line emission
extend east-west, the X-ray emission is wedge-shaped, opening 
toward the south (Paper I). 
Mrk 463 is similar to Mrk 78, containing
a small  ($1\arcsec\ = 1$ kpc) linear radio structure
\citep{Ung86,Maz91,Kuk99} that is associated with [\ion{O}{3}]
line emission \citep{Uom93}. 
Mrk 1066 also has a small (300 pc) linear radio jet and
coincident [\ion{O}{3}] emission \citep{Bow95}. 
The northern X-ray structure
of Mrk 266 has been called a ``jet,'' but there is no definitive 
evidence (from kinematics or additional observations at other wavelengths) 
for this physical origin.  Furthermore,
even morphologically this identification is suspect, for the
emission is not strongly collimated.
No AGN-driven outflows have been conclusively identified in any 
of the other sample members.  

The absence of resolved X-ray emission in most ordinary Seyfert 2
galaxies further argues against its AGN origin in the composite objects.
Where small-scale radio jets are observed,  the
X-ray emission is not necessarily resolved 
(e.g., Mrk 3; \citealt{Kuk93}, \citealt{Mor95}). 
No systematic studies of extended X-ray emission in 
Sy 2s exist, but few of the individual examples that have been examined
exhibit the characteristic resolved thermal emission on the
large scales we detect in the Sy2/SBs \citep{Col98}.  
Moreover, several of the exceptional cases that are resolved do have some
starburst characteristics 
(e.g., NGC 4388; \citealt{Mat94}, \citealt{Leh95}). 

The best explanation for this characteristic emission is the 
starbursts, which produce large volumes of hot gas in superwinds.
Nearby starburst galaxies such as M82 and NGC 253 have kpc-sized X-ray 
emission from starburst-driven winds \citep{Dah98}, while the
starburst-driven X-ray halo of NGC 3628 is even larger, extending
25 kpc \citep{Dah96}.
These are comparable to the observed extended X-rays in our sample. 
In the more inclined examples (Mrk 1066 and Mrk 78), the X-ray
halos extend most obviously perpendicular to the planes of the galaxies,
as expected from starburst-driven outflows.
Furthermore,
we have selected the sample to include only galaxies that definitely
contain a starburst, so it is certainly an existing energy source in 
all cases. 

\subsection{X-ray and IR Flux Ratios}
We compare this sample of Sy2/SB composite galaxies with starbursts, 
Seyfert 2s, and Seyfert 1s in order to identify empirical diagnostics
of their complex nature.  Because the focus of the present work
is the composite galaxies, we rely on published comparison samples.
These comparison galaxies 
have not been selected in the same rigorous way as the
Sy2/SB sample.  The X-ray measurements of
Sy 2 and SB galaxies are particularly
sensitive to modelling, and we
use data that have been analyzed similarly for these
two groups.  The Sy 1s are least sensitive to modelling because
the intrinsic AGN dominates the X-ray emission across
the bandpass we examine.

The starbursts consist of the
far-infrared flux-limited sample of edge-on galaxies of \citet{Dah98} 
and the corresponding face-on 
sample of D. K. Strickland, K. A. Weaver, \& T. M. Heckman (in preparation).
The Sy 2s are taken from \citet{Wea00}, which consists of those
Seyfert 2 galaxies for which archival \asca{} observations were available
as of 1996 August.  This group 
includes some Sy 2 galaxies that contain starbursts 
but which
are excluded from the present composite sample because
they were not compiled in 
the original \citep{Whi92} list from which 
\citet{Hec95} drew their sample,
or because they failed to meet the latter's flux criteria.
In subsequent figures, we identify these galaxies
(NGC 1667, NGC 1808, NGC 4945, and Circinus) 
with both the ``Sy 2'' and ``Sy2/SB'' symbols.
The Sy 1 comparison sample consists of those 
galaxies compiled by \citet{Mas95} that have measurements of 
both hard and soft X-ray luminosities.
Because of their sample selections, the Sy 2 and Sy 1 groups risk
favoring of X-ray-bright sources.  With the combination
of luminosity ratios and absolute luminosities that we explore,
however, this does not appear to be a significant bias.  

The X-ray luminosities we consider are in the soft (0.5--2 keV) and 
hard (2--10 keV) bands, $SX$ and $HX$, respectively.  We correct
these for Galactic absorption alone.  Note that $HX$ is {\em not}
corrected for the measured intrinsic absorption of each source.
The far-infrared emission provides another basis for comparison. 
The FIR flux is computed from $f_{60}$ and $f_{100}$, 
the IRAS 60\micron\ and 100\micron\ flux densities, respectively, measured
in Jy, to account for all far-infrared emission;
$F_{FIR} = 
         1.26\times10^{-23}(2.58\times10^{12}f_{60}+1.0\times10^{12}f_{100})$.
\begin{center}
\includegraphics[width=3.5in]{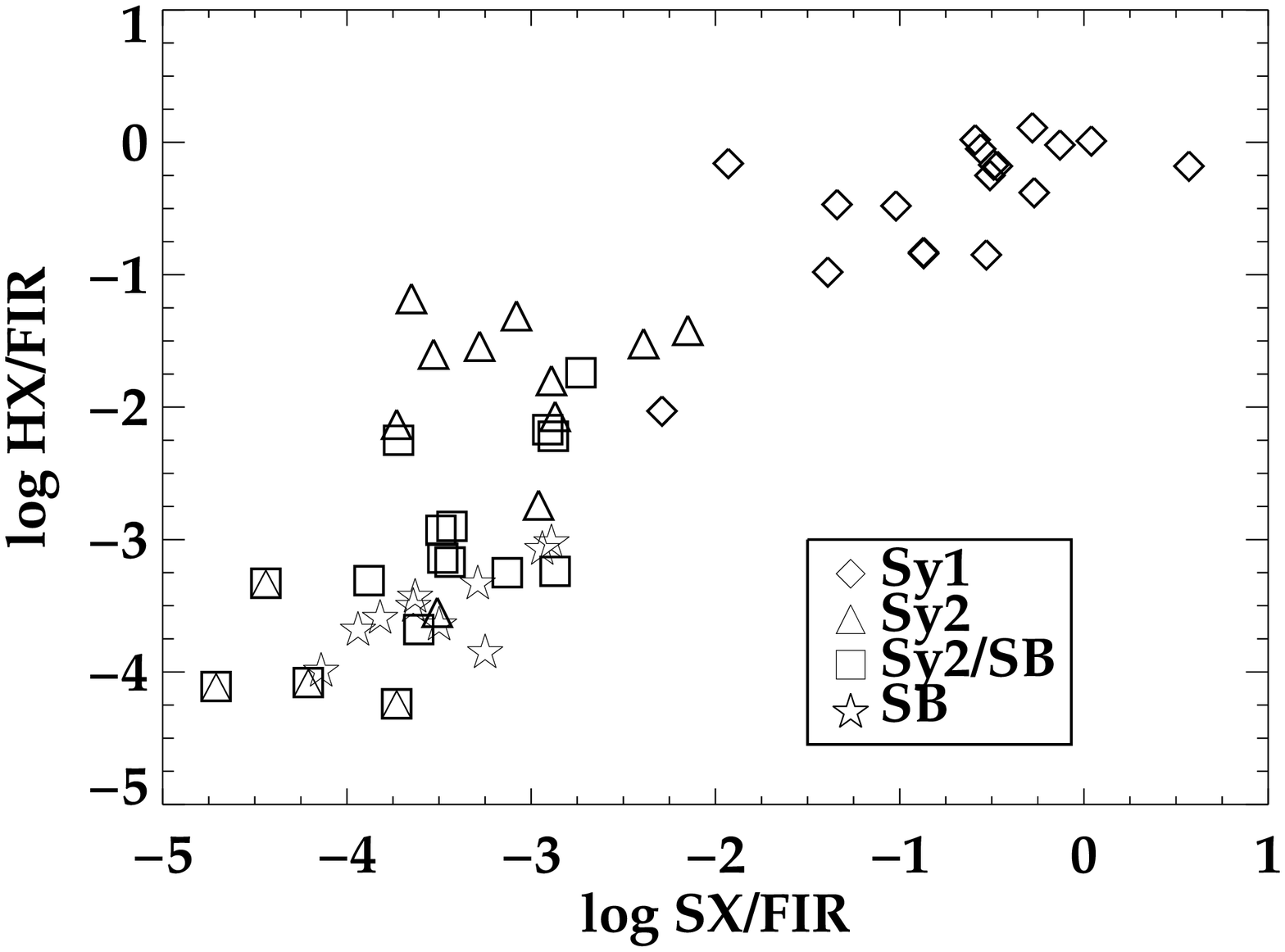}
\end{center}
\vskip -0.2in
\figcaption
{Hard (2.0-10 keV) vs. soft (0.5-2.0 keV) X-ray luminosity, normalized
by FIR luminosity for the Sy2/SB sample ($\Box$), other Sy 2 galaxies ($\triangle$; Weaver et al. 2000), 
starburst galaxies ($\star$; Dahlem et al. 1998, 
D. K. Strickland, K. A. Weaver, \& T. M. Heckman, in preparation),  
and Sy 1 galaxies ($\diamond$; Mas-Hesse et al. 1995).
True composites in the Weaver et al. (2000) sample are plotted with both Sy 2 and
Sy2/SB symbols. 
The composite sample members have $SX/FIR$ ratios similar to
starburst galaxies and $HX/FIR$ in between ordinary Sy 2 and starburst
galaxies.  The Sy 2 examples having relatively low $HX/FIR$ either 
exhibit significant star formation or are heavily absorbed.
\label{fig:sxfirvhxfir}
}
\medskip

In terms of $FIR$, the far-infrared luminosity,
the ratio of $SX/FIR$ in the Sy2/SB galaxies is 
similar to that of ordinary starburst galaxies 
but distinct from those of ordinary Seyfert 2 galaxies.
The $HX/FIR$ ratio further separates the Sy2/SBs from Seyfert 2s
(Figure \ref{fig:sxfirvhxfir}).
As expected, the Sy 2 galaxies of the \citet{Wea00} sample that
are known to contain starbursts 
are spectrally similar to our flux-limited Sy2/SB sample.
The composites in which the thermal component is relatively weak
are most like other AGN.  These sources have a high $HX$, either
due to low absorption, high intrinsic AGN luminosity, or both.
\begin{center}
\includegraphics[width=3.5in]{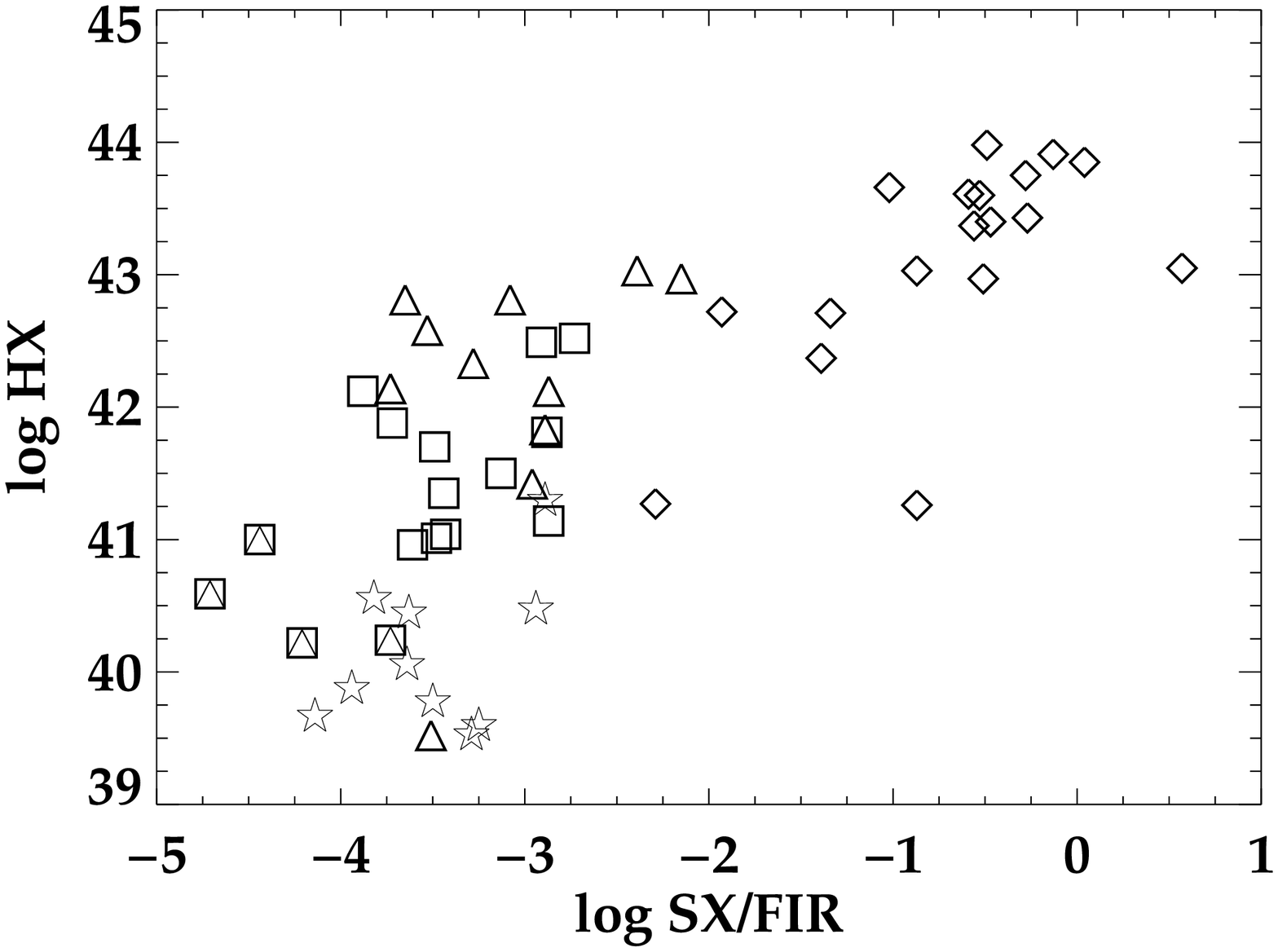}
\end{center}
\vskip -0.2in
\figcaption{The higher observed hard X-ray (2.0-10 keV) luminosities of the Sy2/SB 
sample distinguish
them from starburst galaxies.  The strong absorption of nearly all of the
composites may account for their relatively low $HX$ with respect
to the other Sy 2 galaxies.  (Symbols as in Figure \ref{fig:sxfirvhxfir}.)
\label{fig:sxfirvhx}
}
\medskip

The hard X-ray luminosity distinguishes the
starburst and composite galaxies (Figure \ref{fig:sxfirvhx}). 
A Kolmogorov-Smirnov test demonstrates that $HX$
of the  Sy2/SB and SB groups is different
to a confidence level greater than 99.9\%.
We find a maximum hard X-ray luminosity of
about $10^{41} {\rm \, erg\, s^{-1}}$ for starburst galaxies, while the composites fall
between starbursts and Seyfert 2s.
These two groups are spectrally similar 
\citep[e.g.,][]{Dah98}, 
with X-ray binaries, rather than an active nucleus, 
generating the power-law emission of the starbursts.
The central concentration of the X-ray emission, higher
luminosities, and the Fe K$\alpha$ lines of the Sy2/SB examples imply that 
starburst phenomena alone are not the sole source of X-rays in these
objects.  Each genuinely contains an active nucleus, as well.

The high X-ray luminosity in both hard and soft bands 
distinguish the Sy 1s from all the other
types we consider.  We find $\log (SX/FIR) \gtrsim -2$,
$\log (HX) \gtrsim 42.5$, and $\log (HX/FIR) \gtrsim -1$ in the Sy 1s.
Comparing $SX$ with the $12\micron$ luminosity, $L12$, we find 
that the ratio $SX/L12$ 
does not distinguish the composites from the pure AGNs, but it is
a good discriminant of Sy 1s and Sy 2s.  In agreement with 
Rush et al. (1996)\nocite{Rus96}, all the sources 
having $\log (SX/L12) > -1 $ are Sy 1s, while 
the active galaxies with $\log (SX/L12) < -2 $ tend to be Sy 2s.
The ratio  $SX/L12$ we observe over the sample of starbursts 
is similar to the Sy 2s, so this ratio alone is unlikely to be a reliable
indicator of the presence of an AGN.

The shapes of the Sy2/SB X-ray spectra strongly depend 
on the amount of intrinsic absorption,
which varies widely across the sample.  
When $N_H$ is small,  we see the AGN shining 
through (e.g., NGC 7582), while for the heavily absorbed cases 
(e.g., NGC 5135) the 
galaxies dominate the spectra.    
The intrinsic hard X-ray luminosities are sensitive to the models
because they depend entirely on correcting for absorption,
which is uncertain, especially when $N_H$ is large.  Thus, it is difficult to 
measure the intrinsic physical properties of the AGN.
We determine the absorption-corrected
hard X-ray luminosity, $L_{2-10}^{ac}$, to compare our results
with others, but we base our conclusions above on the {\em observed}
hard X-ray luminosity from 2 to 10 keV, $HX$.  

The observed hard X-ray luminosity in the Sy2/SBs is somewhat lower than $HX$
measured in other Seyfert 2 galaxies, and the 
the heavily-absorbed composites are significantly 
underluminous in observed hard X-rays
with respect to Seyfert 1s.
If the unified model is correct, 
in which the Seyfert classification is determined merely by orientation, then
we expect Seyfert 2s to have intrinsic hard X-ray properties
similar to Seyfert 1s.
After correcting for the largest amount of absorption that we can
measure with \asca, however, we find that $L_{2-10}^{ac}$
in the composite sample is systematically lower than the hard X-ray luminosity
of Sy 1 galaxies.  This conclusion holds true even when we ignore
the Compton thick sources, those in which we do not observe the AGN directly.
Awaki et al. (2000)\nocite{Awa00} suggest 
that polarized broad line Seyferts are underluminous because they are
completely blocked.  In other words, the observed spectrum is a
complex mix of reflected and then absorbed flux.  In this case, the
true hard X-ray luminosity would be a factor of about 10 higher than the intrinsic flux
derived purely from an absorption correction.  This would require
multiple regions of matter to attenuate and scatter the intrinsic
power-law spectrum.  If $L_{2-10}^{ac}$ of the composite sample
is increased tenfold, the hard X-ray fluxes do become 
comparable to those of Sy 1s.
\begin{center}
\includegraphics[width=3.5in]{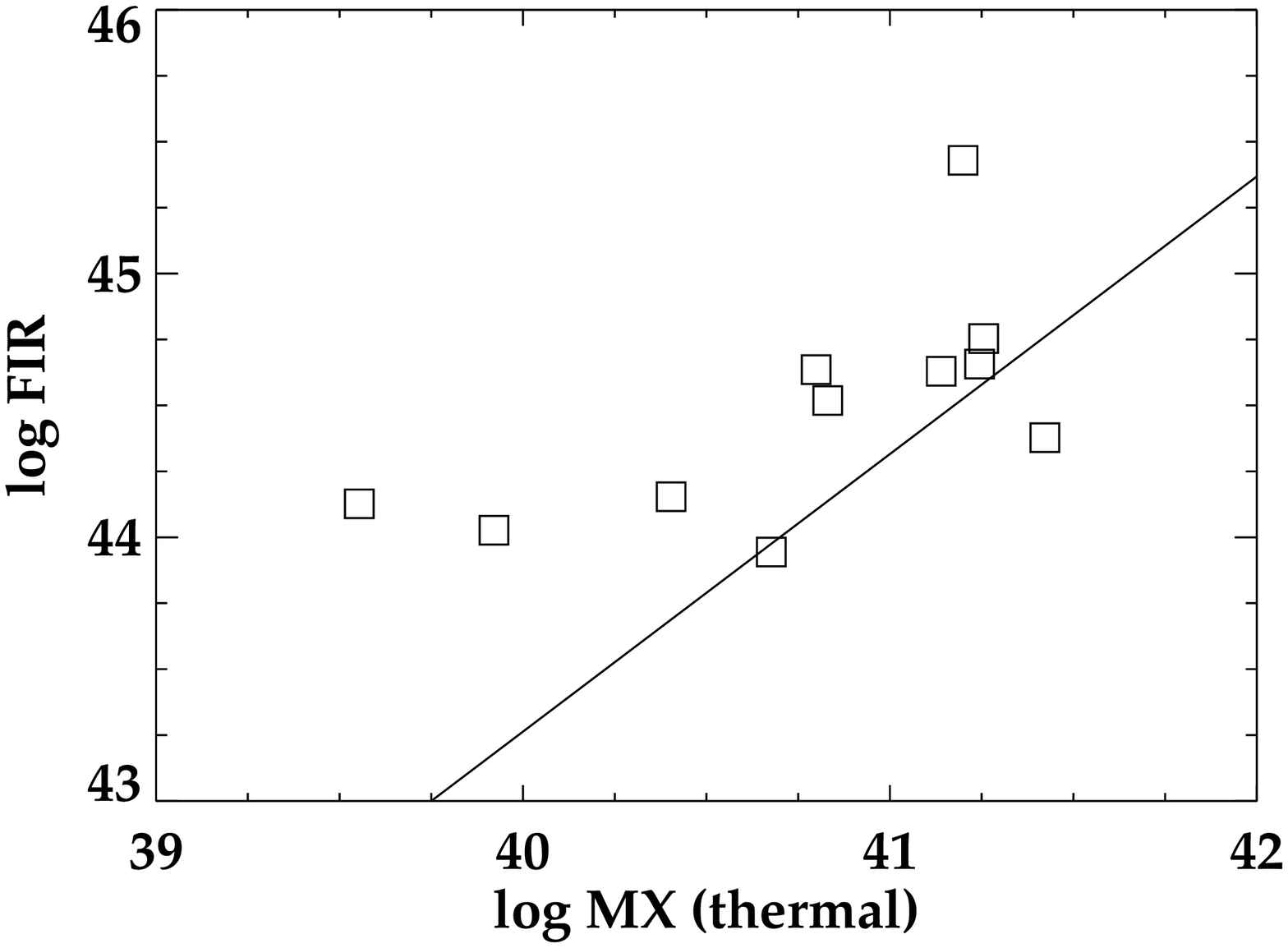}
\end{center}
\vskip -0.2in
\figcaption{Thermal component of medium (0.5-4.5 keV) X-ray luminosity vs.
FIR luminosity for Sy2/SB sample ($\Box$), corrected for Galactic absorption.  
The correlation between thermal X-ray and FIR luminosities observed
in the composite galaxies agrees with the empirical correlation for normal
and starburst galaxies ({\it solid line}; \citealt{Dav92}).
This consistency implies that both are related to star formation 
processes and are not due to AGNs. 
\label{fig:firmxt}
}
\medskip

\citet*{Dav92} 
determined an empirical relation between the 
medium X-ray (0.5--4.5 keV) and FIR luminosity for normal
and starburst galaxies.  The observations of our sample
are consistent with this relation.  In particular, the FIR
luminosity of our sample members follows
this relation when {\em thermal} X-ray luminosity
alone---the starburst component---is considered (Figure \ref{fig:firmxt}).  
This suggests that similar to normal galaxies, the FIR luminosity 
is related to the stellar activity in the Sy2/SBs.
Reprocessed AGN light also adds somewhat to FIR emission, so $FIR$ in
the composite galaxies tends to be slightly higher than that associated
with the thermal X-ray component alone.
For a given FIR luminosity, the 
{\em total} medium X-ray luminosity of the Sy2/SBs is 
consistently greater than expected from the normal and starburst
galaxy correlation by factors of several 
because of the additional AGN contribution to $MX$.
The composites are similar to normal Sy 2s with their high 
$MX_{total}/FIR$ ratios compared with starburst galaxies 
\citep[e.g.,][]{Gre92}.
Thus, these data imply that similar to normal and starburst galaxies, 
star formation predominantly determines FIR emission, 
and the thermal X-ray emission traces this star-forming component.

\subsection{Column Density and Starbursts}
The strength of the fluorescent Fe K$\alpha$ line is a function of 
column density.  
In the Sy2/SBs, we find that the Fe line
equivalent widths (EWs) are consistent with simple geometrical
scattering models, although we cannot discriminate between them
in detail (Paper I, Figure 12).  
This result requires accurately accounting for intrinsic and
scattered continuum flux to identify the large column densities
of this sample.
\begin{center}
\includegraphics[width=3.5in]{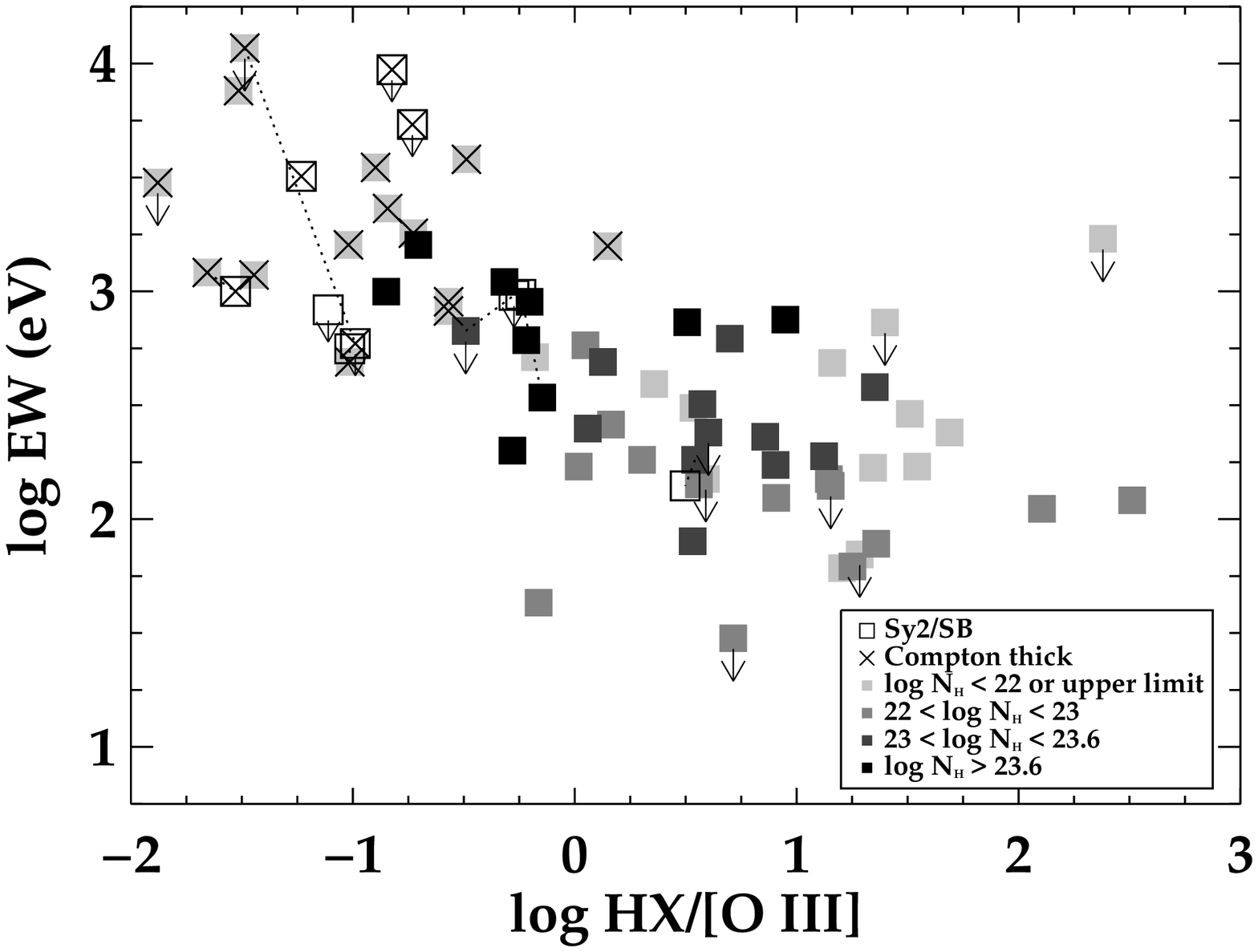}
\end{center}
\vskip -0.2in
\figcaption
{Fe K equivalent width vs. hard X-ray/[\ion{O}{3}] flux ratio for
Sy 2s compiled by Bassani et al.\ (1999) 
and our sample of composite galaxies.
Measurements of the same source in both works are connected
with a dotted line.
A large obscuring column suppresses the observed hard X-rays
without altering the [\ion{O}{3}] flux.  The EW is measured with respect to 
this diminished continuum and is thus relatively large.  These
effects together help to distinguish the Compton thick sources.
\label{fig:ewhxo3}}
\medskip

As Bassani et al. (1999)\nocite{Bas99} demonstrate empirically for
a sample of Sy 2s, 
the EW combined with the ratio of hard X-ray to 
[\ion{O}{3}] emission is an
indicator of $N_H$.  The hard X-ray flux is diminished
by the intervening column density, while the isotropic 
[\ion{O}{3}] emission is not directly affected.  
Increasing $N_H$ thus tends to decrease the ratio $HX/$[\ion{O}{3}].
In the Compton thick regime ($N_H > 10^{24}\psc$), the emerging 
continuum at 6.4 keV is suppressed, so the measured EW is therefore
relatively large.  Indeed, plotting the Sy2/SBs with the
Bassani et al. (1999)\nocite{Bas99} sample, further illustrates
their large column densities (Figure \ref{fig:ewhxo3}).

The distribution of absorbing column density in Sy2/SB composites 
is significantly different from that measured in ``pure'' Sy 2s.
We compare our sample with the [\ion{O}{3}] flux-limited Sy 2
sample of \citet*{Ris99},
excluding known composites.
The two samples are different at the 99\% confidence level.
The column densities of the Sy 2s with no evidence of a 
starburst are distributed over
a wide range, with 27\% having $N_H < 10^{23}\psc$. 
In contrast, the Sy2/SBs tend to have higher directly measured or inferred 
column densities, with $N_H < 10^{23}\psc$ only in the unusual case of 
NGC 6221.  This result is not a selection effect.  
The initial sample was flux-limited, based on isotropic
AGN properties, and low column densities in these Sy 2s would not
affect the optical or UV detectability of the starburst. 

The nuclear column densities of molecular gas in 
ordinary starburst galaxies 
range from $10^{22}$ to $10^{25} \psc$ (e.g., 
\citealt{Dah98}, and \citealt{Ken98}), 
and in ultraluminous 
galaxies, $N_H\sim 10^{23}$ to $10^{25}\psc$ \citep{Gen98}.
These column densities are comparable to those measured toward 
the central engines of Sy 2s.
The presence of a dense ISM on circumnuclear scales may explain
why the hard nuclear source is generally heavily obscured
in the composite objects.  Based on the new results from our 
Sy2/SB sample, we propose that 
the high inferred X-ray column densities
are not solely due to the obscuring torus in these galaxies.
{\em The circumnuclear starburst also obscures the AGN.}
\begin{center}
\includegraphics[width=3.5in]{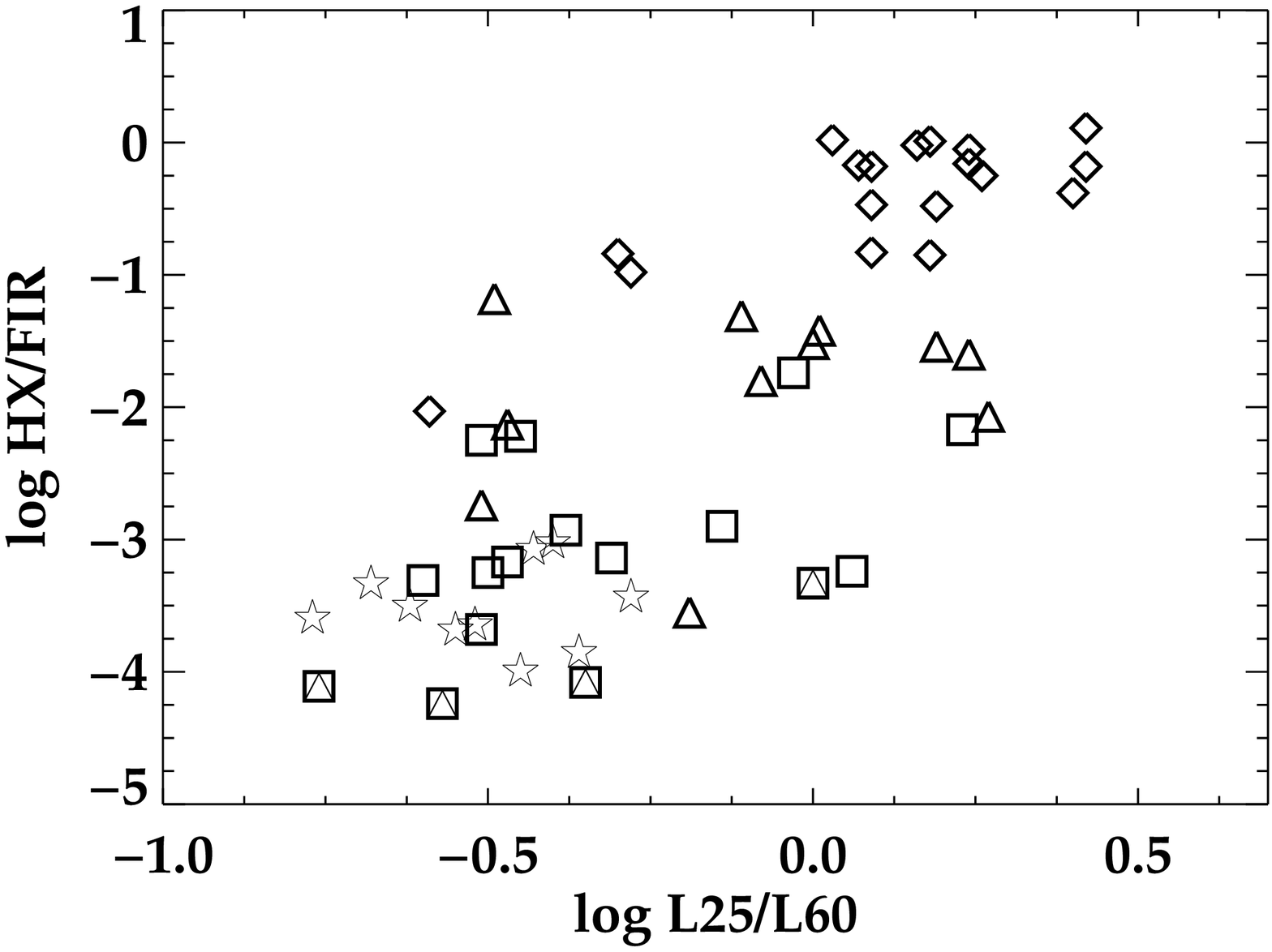}
\end{center}
\vskip -0.2in
\figcaption
{Hard X-ray to far-infrared luminosity vs. 25 to 60 \micron\ luminosity
for Sy2/SBs and comparison samples.  The far-infrared luminosity ratio
indicates the relative strength of the AGN and starburst components,
with higher values observed when the AGN dominates.  The ratio
$HX/FIR$ decreases with obscuration.  The strong correlation between
these two ratios and the absence of many objects from the 
high-$L25/L60$, low-$HX/FIR$ regime (lower right corner) supports
the suggestion that the starburst itself both alters $L25/L60$
and obscures the AGN in composite objects.
(Symbols as in Figure \ref{fig:sxfirvhxfir}.)
\label{fig:2560vhxfir}
}
\medskip

Comparing the ratio of luminosity at 25 and 60 \micron\ ($L25/L60$) with
$HX/FIR$ illustrates the association
of starburst activity and obscuration (Figure \ref{fig:2560vhxfir}).  
The ratio $L25/L60$ indicates relative strength of 
AGN activity compared with star formation, while 
$HX/FIR$ is a function of obscuration.
With increasing obscuration,
hard X-rays are blocked, so $HX$ decreases.
More material is also available to reprocess the
intrinsic emission to the far-infrared band, so $FIR$ increases. 
As the Sy 1s show, relatively high values of both of these ratios
indicate a relatively strong AGN {\em and} low obscuration.
Including a starburst in an active galaxy not only
changes the FIR color, 
but it also tends to obscure the AGN. 
Most significantly, few galaxies exhibit high values of $L25/L60$
and low $HX/FIR$ simultaneously.  Thus, obscuration of the central 
engine seems to correlate with the presence of energetically significant
star formation.  While the comparison samples were not rigorously
selected, this correlation is not due to 
selection effects; heavily obscured and AGN-strong
candidates are not preferentially excluded from the comparison groups.

Luminous infrared galaxies ($FIR > 10^{11} L_{\sun}$) 
exhibit a similar distribution of these flux ratios.
To explain these observations,
Risaliti et al. (2000)\nocite{Ris00} propose that the infrared and X-ray
emission suffer different absorption in these galaxies because of 
a particular viewing geometry.  
Our results support the suggestion that the X-ray source is
more obscured, but it is not a function of viewing angle with
respect to the circumnuclear torus.  Rather, we 
specifically identify the starburst, the source of strong 60\micron\ emission, 
as some of the additional medium that blocks the AGN.
A compact torus may still be present, and its orientation may
fundamentally account for
the classification of these galaxies as Sy 2s rather than Sy 1s,
but the torus is not the only absorber.

Observations of bars in Sy 2 galaxies provide further indirect 
evidence that column density is related to
the presence of a starburst.
The column density in barred Sy 2s is significantly greater than
in non-barred Sy 2s, and more than 80\% of the Compton thick
Sy 2s are barred \citep{Mai99}.
In general, stellar bars are strongly associated with circumnuclear 
star formation \citep{Ken98}.  Thus, Sy 2s with bars are likely to 
have high central star formation rates, which we associate 
with high column density.  The physical sizes of obscuration
are appropriate, with bar lengths around 100 pc, which is similar to the
starburst scales.

The presence of a circumnuclear starburst is a good indicator
of large column density.  This is true not only in our sample 
but also in larger samples with different selection criteria, 
such as \citet{Ris99}.
In this other sample, we find that many of the heavily
obscured active galaxies also contain significant star formation.
Because we attribute some of the obscuration to the starburst directly, 
we suggest that the converse may also hold, that large column density 
may be a starburst indicator.  
Indeed, all the Compton thick galaxies of the \citet{Ris99} sample
in which the circumnuclear stellar populations have been studied in detail
do contain starbursts.
Specifically, the column density is measured in 36 members of this
sample, and 14 of these are Compton thick.  Eight of these
14 have been examined for evidence of starbursts, and all
eight show evidence for starbursts.
The possible composite nature of the remaining 6 cases is unknown because
they have not been examined for indications of star formation.

\subsection{The Role of Galaxy Interactions}

With the exceptions of Mrk 1066 and Mrk 78,  
all these Sy2/SB composites show evidence for galaxy interactions or mergers.  
They either contain 
multiple nuclei, show evidence for interaction with companion 
galaxies, or are members of groups or clusters.  
The high incidence of interactions implies that this 
phenomenon relates to the starburst, probably triggering the 
star formation, and is therefore important in efficiently providing 
material to the black hole, either directly---through dynamic 
instabilities---or indirectly---as a consequence of the starburst 
winds and supernovae.  
While the sample numbers are too small to provide statistically
significant results, we note that only four of the seven Seyfert
2 galaxies {\em without} starbursts
from the original optical and radio flux-limited sample show
evidence for galaxy interactions or mergers or are located in clusters
or groups. 
Starbursts may be important in the complete
cycle of development of luminous active galaxies, possibly including 
the initial creation of the central black hole \citep{Nor88},
through the clearing of obscuring material to allow lines of sight
to the AGN directly in type 1 objects at later times \citep{Hec89}.

Recent work \citep{Lau95,Dul99} shows that Seyfert 2 galaxies 
tend to have more companions than comparison samples of
Seyfert 1s and normal galaxies.
(In these
studies, companions are ``nearby'' projected onto the plane of the
sky, not necessarily physically close.) 
In particular, Laurikainen \& Salo (1995)\nocite{Lau95} 
find that the Sy 2s with
companions tend to be morphologically peculiar galaxies.  
These are 
most likely to contain starbursts.
Seyfert 1s are not preferentially located in denser environments.
Because the environment is independent of viewing angle of the
AGN and surrounding material, this result challenges the 
unified model.  
As we suggest above, the starbursts themselves may literally obscure 
the unity of the two classes. 
Interacting galaxies may generate circumnuclear starbursts that obscure the
central engine and prevent direct observation of a broad line region.
Thus, interacting Seyfert galaxies, a subset of those with companions as
defined in large-scale surveys, 
are more likely to be type 2 than type 1.

\section{Summary and Conclusions\label{sec:concl}}

In this work, we analyze a selection of Seyfert/starburst composite galaxies.
These are active galaxies that also contain significant star formation.
The sample is flux-limited, based on optical 
([\ion{O}{3}] $\lambda\lambda 4959+5007$) or radio (1.4 GHz) nuclear fluxes,
and the starbursts are detected in optical and UV observations.

The results obtained from both X-ray imaging and spectroscopy agree.
In the imaging data, we measure fractions of resolved emission that 
are comparable to the thermal proportions identified in the spectral models.
Thus, association of the spatially extended and spectrally thermal 
emission---common characteristics of starbursts in X-rays---is appropriate. 
The thermal component generally accounts for a significant fraction of
the total soft X-ray emission in these galaxies.  The spectra are
genuinely complex, and simple models, such as a single power law,  
that attempt to characterize the intrinsic AGN properties are inapplicable
and yield misleading results.

With one exception (NGC 6221), all the AGNs are heavily-absorbed,
with $N_H>10^{23}\psc$,
and the large equivalent widths of the Fe K$\alpha$ 
lines are consistent with these high column densities.
The column densities of this 
sample are significantly different from a comparison ``pure''
Sy 2 sample.  Some fraction of the absorber is likely
the starburst itself.  We propose that highly obscured AGNs
may preferentially contain starbursts. 
We do not pursue comparisons with Sy 1s because the corrections for
intrinsic properties are highly uncertain, but we do note that 
the total hard X-ray luminosities of the Sy2/SBs are systematically lower
than $HX$ of the comparison Sy 1s, even after simple
correction for the measured absorption.

The best broad-band discriminant of Seyfert/starburst 
composites in general samples of 
galaxies is the ratio of soft X-ray to far-infrared emission,
$SX/FIR$, combined with a hard X-ray measurement.  
Where it is significant, star formation is ultimately 
directly tied to much of the soft X-ray
and FIR luminosity.  These flux ratios of the Sy2/SBs are similar
to those of ordinary starburst galaxies, while the AGN produces
the hard X-rays.  In contrast, pure Sy 2s are more similar to
Sy 1s, where nearly all X-ray emission is due to the AGN.

Most of these composite galaxies show signs of interactions or mergers,
such as
multiple nuclei and membership in groups or clusters
demonstrates.  This dynamic activity probably
triggers star formation, possibly as part of the 
evolutionary relationship between starburst and Seyfert galaxies. 
As Heckman et al. (1989)\nocite{Hec89} describe, a starburst
phase may be one stage in the development of an AGN. 

\begin{acknowledgements} 
This work was supported by NASA grants  NAG5-6917 and NAG5-6400.
This research has made use of the NASA/IPAC Extragalactic Database
(NED) which is operated by the Jet Propulsion Laboratory,
California Institute of Technology, under contract with the
National Aeronautics and Space Administration,
the Astronomical Data Center at NASA Goddard Space Flight Center,
and the High Energy Astrophysics Science Archive Research Center Online Service
provided by the NASA Goddard Space Flight Center. 

\end{acknowledgements} 

\clearpage 

\rotate
\thispagestyle{empty}
\begin{deluxetable}{lllllllrrcccllclc}
\tablewidth{0pt}
\tablecaption{Basic Properties of the Selected Galaxies\label{tab:infoir}}
\tablehead{
\colhead{Galaxy}&\multicolumn{3}{c}{RA}&\multicolumn{3}{c}{DEC}&\colhead{D} &\colhead{Scale}
&\colhead{Galactic $N_H$}
&\colhead{$F_{12}$}&\colhead{$F_{25}$}&\colhead{$F_{60}$}&\colhead{$F_{100}$}
&\colhead{$i$}& 
\colhead{References}\cr
&\colhead{(h}&\colhead{m}&\colhead{s)}&\colhead{($^\circ$}&\colhead{\arcmin}&\colhead{\arcsec)}
&\colhead{(Mpc)}&\colhead{(pc/$\arcsec$)}&\colhead{($10^{20}{\rm \,cm^{-2}}$)}&
\colhead{(Jy)}&\colhead{(Jy)}&\colhead{(Jy)}&\colhead{(Jy)}&\colhead{($^\circ$)}
}
\startdata
NGC 1068  & 02& 42& 40.71   & $-00$&  00&  47.8 & 15.2  &  74 & 3.0 &39.7 & 85.0 & 176 &  224 & 32 &1, 4\\   
Mrk 1066  & 02 & 59 & 58.59 & $+36$ & 49 & 14.3 & 48  & 233 & 12  &0.45 & 2.26 & 11.0 &  12.1 & 54 &2, 4, 6 \\
Mrk 1073  & 03 & 15 & 01.43 & $+42$ & 02 & 09.4 & 93  & 452 & 1.1 &0.44 & 1.41  & 8.17 &  11.1 &42 &2, 5, 6 \\
Mrk 78    & 07 & 42 & 41.73 & $+65$ & 10 & 37.5 & 149 & 720 & 4.1 &0.13 & 0.55  & 1.11 &  1.13 & 60 &2, 4, 6\\
IC 3639   & 12 & 40 & 52.88 & $-36$ & 45 & 21.5  & 44  & 212 & 5.1 &0.64 & 2.26 & 7.52 &  10.7 & 34 &2, 5, 7\\
NGC 5135  & 13 & 25 & 43.97 & $-29$ & 50 & 02.3  & 55  & 266 & 4.7 &0.64 & 2.40 & 16.9 &  28.6 & 46 &2, 4, 7, 8, 9\\
Mrk 266   & 13 & 38 & 17.69 & $+48$ & 16 & 33.9  & 111 & 540 & 1.5 &0.23 & 0.98 & 7.34 &  11.1 & \nodata & 1, 4, 10\\
Mrk 273   & 13 & 44 & 42.11 & $+55$ & 53 & 12.6 & 151  & 732 & 0.97 &0.24  & 2.28 & 21.7 &  21.4 & \nodata & 1, 5, 6\\
Mrk 463   & 13 & 56 & 02.87 & $+18$ & 22 & 19.5 & 199 & 963 & 2.1 &0.51 & 1.58  & 2.18 &  1.92 & 10 & 2, 5, 6\\
Mrk 477   & 14 & 40 & 38.11 & $+53$ & 30 & 16.0 & 151 & 733 & 1.3 &0.13 & 0.51  & 1.31 &  1.85 &\nodata & 2, 5, 7, 11\\
NGC 6221  & 16 & 52 & 46.67 & $-59$ & 12 & 59.0  & 20  &  96 & 15 &1.49  & 5.27 & 36.3 &  84.5 & 43  & 2, 5, 8\\
NGC 7130  & 21 & 48 & 19.48 & $-34$ & 57 & 09.2 & 65  & 313 & 2.0 &0.59 & 2.12  & 16.5 &  25.6 & 29 & 2, 5, 7, 8, 9\\
NGC 7582  & 23 & 18 & 23.50 & $-42$ & 22 & 14.0 & 21  & 102 & 1.5 &1.62 & 6.44  & 49.1 &  72.9 & 62 &3, 5, 8, 9 \\
NGC 7674  & 23 & 27 & 56.72 & $+08$ & 46 & 44.5 & 116 & 563 & 5.2 &0.67 & 1.90  & 5.59 &  8.15 & 25 &2, 4, 6\\
\enddata

\tablerefs{
(1) Murphy et al. 1996; 
(2) Dickey \& Lockman 1990; 
(3) Elvis, Lockman, \& Wilkes 1989; 
(4) Schmitt et al. 1997; 
(5) Whittle 1992;        
(6) Gonz\'alez-Delgado et al. 2000; 
(7) Gonz\'alez-Delgado et al. 1998; 
(8) Cid Fernandes et al. 1998; 
(9) Schmitt et al. 1999; 
(10) Wang et al. 1997; 
(11) Heckman et al. 1997 
}
\end{deluxetable}

\begin{deluxetable}{lrllcrllcl}
\tablewidth{0pt}
\tablecaption{Extended and Thermal Soft X-ray Emission\label{tab:extend}}
\tablehead{
&\multicolumn{3}{c}{HRI}&&\multicolumn{3}{c}{PSPC}&&\multicolumn{1}{c}{ASCA}\\
\cline{2-4}\cline{6-8}\cline{10-10}\\ 
\colhead{Galaxy}&\colhead{$R$\tablenotemark{a}$_{kpc}$}&\colhead{$f$\tablenotemark{b}}
&\colhead{$f$\tablenotemark{c}}
&&\colhead{$R$\tablenotemark{a}$_{kpc}$}&\colhead{$f$\tablenotemark{b}}
&\colhead{$f$\tablenotemark{c}}&&\colhead{$f$\tablenotemark{d}$_{therm}$}
}
\startdata
NGC 1068  &  7.4 & $0.51^{+0.05}_{-0.09}$& $0.73 \pm 0.02$      &&  7.4    & $ 0.40   \pm0.02        $ & $ 0.60\pm 0.06$       &&  0.80  \\
Mrk 1066  &  0.0 & $<0.15               $& $<0.71        $      && \nodata &  \nodata                  & \nodata               &&  0.51  \\
Mrk 1073  &  9.5 & $0.53^{+0.18}_{-0.28}$&$0.84^{+0.16}_{-0.50}$&& \nodata &  \nodata                  & \nodata               &&  0.91  \\
Mrk 78    &  17  & $0.38^{+0.20}_{-0.30}$&$0.72^{+0.28}_{-0.54}$&& 36      & $ <0.62                 $ & $0.72^{+0.28}_{-0.45}$&&\nodata  \\
IC 3639   &  6.8 & $0.39^{+0.11}_{-0.21}$& $0.68 \pm 0.31$      && \nodata &  \nodata                  & \nodata               &&  1.00  \\
NGC 5135  &  5.3 & $0.46^{+0.06}_{-0.16}$& $0.65 \pm 0.20$      &&  13     & $ 0.19  ^{+0.07}_{-0.11}$ & $ < 0.32      $       &&  0.54  \\
Mrk 266   &  22  & $0.78^{+0.06}_{-0.16}$&$0.92^{+0.08}_{-0.20}$&&  32     & $ 0.49  ^{+0.08}_{-0.12}$ & $0.77^{+0.23}_{-0.28}$&&  0.43  \\
Mrk 273   &  18  & $<0.39               $&$0.53^{+0.47}_{-0.49}$&& 37      & $<0.28                  $ & $ < 0.62      $       &&  0.43  \\
Mrk 463   & \nodata            & \nodata &   \nodata            && 48      & $<0.41                  $ & $ < 0.67      $       &&  0.31  \\
Mrk 477   &  15  & $<0.30               $& $0.54 \pm 0.35$      && \nodata &  \nodata                   & \nodata              &&  0.00  \\
NGC 6221  &  4.8 & $0.52^{+0.07}_{-0.17}$&$0.96^{+0.04}_{-0.22}$&& \nodata &  \nodata                   & \nodata              &&  0.05  \\
NGC 7130  &  16  & $0.44^{+0.23}_{-0.33}$& $< 0.63       $      && \nodata &  \nodata                   & \nodata              &&  0.57  \\
NGC 7582  &  3.1 & $0.56^{+0.09}_{-0.19}$&$0.94^{+0.06}_{-0.27}$&&  5.1    & $ 0.38  ^{+0.08}_{-0.12}$ & $ 0.54\pm 0.28$       &&0.16, 0.28  \\
NGC 7674  & \nodata            & \nodata &   \nodata            && 45      & $0.65   ^{+0.48}_{-0.52}$ & $ < 0.57      $       &&\nodata  \\
\enddata
\tablenotetext{a}{Maximum radius of extended emission (kpc).}
\tablenotetext{b}{Minimum extended fraction of total emission, with 90\% confidence errors.}
\tablenotetext{c}{Constrained measurement of extended fraction of total emission.} 
\tablenotetext{d}{Thermal fraction of soft emission in spectrum.} 
\end{deluxetable}

\thispagestyle{empty}
\hoffset -1in
\begin{deluxetable}{lllllllclllllc}
\tabletypesize{\scriptsize}
\tablewidth{0pt}
\tablecaption{Best-Fitting Models ($\Gamma=1.9$)\label{tab:best1}}
\tablehead{
&\multicolumn{6}{c}{Hard Component}&&\multicolumn{5}{c}{Soft Component}\\
\cline{2-7} \cline{9-13}\\
\colhead{Galaxy}&\colhead{$N$\tablenotemark{a}$_H$}&\colhead{$A1$\tablenotemark{b}}
&\colhead{$E$\tablenotemark{c}$_{line}$}
&\colhead{$\sigma$\tablenotemark{d}$_{line}$}
&\colhead{$EW$\tablenotemark{e}$_{line}$}
&\colhead{$F$\tablenotemark{f}$_{2-10}$}
&&\colhead{$N$\tablenotemark{a}$_H$}
&\colhead{$A2$\tablenotemark{g}}&\colhead{$kT$\tablenotemark{h}}
&\colhead{$A3$\tablenotemark{i}}
&\colhead{$F$\tablenotemark{j}$_{0.5-2}$}
&\colhead{$\chi^2/$dof}
}
\startdata
\sidehead{{\it ASCA}}
NGC 1068 & \nodata & \nodata &$6.4^{+0.03}_{-0.06}$ & 0.05f & $1000^{+260}_{-630
}$& $49\pm 1.7$ && 3.0f& $9.0^{+0.50}_{-0.56}$& $0.69\pm 0.02$& $18^{+0.39}_{-0.77}$& $64.2\pm 3.2$&  1870/786 \\
         &\nodata & \nodata &$6.6^{+0.07}_{-0.06}$&0.50f&$4200^{+500}_{-630}$&\nodata&&\nodata&\nodata&$1.66\pm0.2$&$7.6^{+1.6}_{-1.3}$&\nodata&\nodata\\
         &\nodata & \nodata &$6.7^{+0.24}_{-0.08}$&0.05f&$350^{+150}_{-190}$&\nodata&&\nodata&\nodata&\nodata&\nodata&\nodata&\nodata\\
Mrk 1066 & $57^{+59}_{-45:}$ & $0.97^{+0.25}_{-0.29}$ & $6.6^{+0.20}_{-0.52}$ & 0.05f & $3200\pm1900$ & $3.6\pm0.47$ && 12f & \nodata & $0.88^{+0.22}_{-0.20}$& $0.42^{+0.14}_{-0.11}$& $1.3\pm0.17$& 74/73 \\
Mrk 1073 & $400^{+390}_{-220}$&$2.2^{+0.89}_{-0.70}$&\nodata&\nodata&\nodata&$4.7^{+0.74}_{-0.85}$&&14f&\nodata&$1.0^{+0.47}_{-0.29}$&$0.70^{+0.82}_{-0.28}$&$1.3\pm0.26$&119/63\\
IC 3639 &$2100^{+3000}_{-1300}$&$3.4^{+ 3.2}_{-1.7}$&\nodata&\nodata&$<830$&$ 4.8\pm1.2$&&5.1f&\nodata&$ 2.3^{+2.9}_{-0.7}$&$1.7\pm0.53$&$1.2\pm0.29$& 48/53 \\
NGC 5135 &$ 12900^{+16000}_{-6800}$&$ 26^{+120}_{-18}$& \nodata&\nodata&$<1100$&$ 6.2^{+1.5}_{-1.6}$&& 4.7f &$ 0.67^{+0.09}_{-0.15}$&$ 0.77^{+0.04 }_{-0.1}$&$ 0.79^{+0.28}_{-0.14}$&$ 2.8^{+0.5}_{-0.2}$& 108/98 \\ 
Mrk 266 & $160^{+330}_{-150}$&$1.2^{+1.1}_{-0.6}$&\nodata&\nodata&$<10000$&$3.1\pm1.1$&&1.5f&\nodata&$0.78\pm0.3$&$0.73^{+0.35}_{-0.39}$&$1.8\pm0.6$&13/33\\
Mrk 273 & $3300^{+2900}_{-1100}$& $3.9^{+2.3}_{-1.9}$& $6.5^{+0.21}_{-0.17}$ & 0.05f&$860^{+130}_{-230}$&$5.2\pm1.0$&& 1.0f &$0.48^{+0.14}_{-0.21}$&$1.1^{+0.88}_{-0.77}$&$0.19^{+0.41}_{-0.17}$ &$1.2\pm0.24$& 55/57 \\ 
Mrk 463 &$ 2900^{+1700}_{-1200}$&$ 6.0^{+4.7}_{-2.8}$& \nodata&\nodata&$<890$&$ 6.6\pm1.3$&&2.1f&$ 0.40^{+0.080}_{-0.086}$&$ 0.71^{+0.53}_{-0.45}$&$ 0.15^{+0.080}_{-0.11}$&$ 1.1\pm0.17$& 80/70 \\
Mrk 477 & $2400^{+1700}_{-1200}$&$8.8^{+8.9}_{-4.3}$&$6.4^{+0.23}_{-0.21}$&0.05f&$560^{+560}_{-500}$&$12^{+2.3}_{-1.9}$&& 1.3f & $0.59^{+0.22}_{-0.20}$& \nodata&\nodata&$1.2\pm0.2$&36/104\\ 
NGC 6221 & $110^{+8.6}_{-8.3}$&$46^{+1.9}_{-1.3}$&$6.6^{+0.31}_{-0.29}$&0.50f&$360^{+210}_{-93}$&$140\pm4.1$&&15f&$3.0^{+1.4}_{-1.5}$&$1.4^{+1.6}_{-0.5}$&$1.2^{+1.7}_{-0.69}$&$24\pm0.7$&799/765\\
NGC 7130 &$ 16^{+62}_{-14:}$&$ 0.60\pm0.3$&\nodata&\nodata&$<5400$&$ 1.8\pm0.3$&&2.0f&\nodata&$ 0.79^{+0.2 }_{-0.1}$&$ 0.56^{+0.29}_{-0.25}$&$ 2.0^{+0.3}_{-0.4}$& 74/74 \\
NGC 7582 (1994) & $1100^{+74}_{-78}$& $84^{+6.6}_{-6.5}$& $6.2^{+0.17}_{-0.33}$ & 0.05f&$190^{+60}_{-140}$&$140\pm7.1$&& 1.5f &$2.0^{+0.33}_{-0.30}$&\nodata&\nodata&$4.4^{+0.6}_{-0.5}$& 344/284\\ 
NGC 7582 (1996)&$1400^{+60}_{-50}$& $93^{+4.6}_{-4.4}$& $6.3\pm0.08$ & 0.05f&$150^{+53}_{-51}$&$140\pm 4.2 $&& 1.5f &$2.1^{+0.14}_{-0.16}$&\nodata&\nodata&$4.4^{+0.2}_{-0.3}$& 516/467\\ 
\sidehead{{\it ASCA + PSPC}}
NGC 1068 & \nodata & \nodata &$6.4\pm 0.05$  & 0.05f & $1000\pm 310$& $50\pm 2.5$ && 3.0f& $11^{+0.27}_{-0.29}$& $0.79\pm 0.01$& $18^{+0.55}_{-0.46}$& $100\pm 5.0$&  2315/910 \\
         &\nodata & \nodata &$6.6\pm0.07$&0.50f&$3500^{+250}_{-390}$&\nodata&&\nodata&\nodata&$0.13\pm0.005$&$110^{+9.2}_{-7.9}$&\nodata&\nodata\\
         &\nodata & \nodata &$6.7^{+0.10}_{-0.14}$&0.05f&$530^{+410}_{-280}$&\nodata&&\nodata&\nodata&\nodata&\nodata&\nodata&\nodata\\
NGC 5135 &$ 14000^{+16000}_{-3700}$&$ 29^{+100}_{-21}$&\nodata&\nodata&$<590$&
$ 6.3\pm0.63$&&4.7f&$ 0.71^{+0.11}_{-0.10}$&$ 0.68^{+0.14}_{-0.06}$&$ 0.79^{+0.10}_{-0.11}$&$ 2.9\pm0.29$& 150/119 \\
Mrk 266 &$1.5:^{+1.1}_{-0:}$&$0.75\pm0.2$ &\nodata&\nodata&$<9400$&$2.1^{+0.4}_{-0.8}$&&1.5f&\nodata&$0.28^{+0.1}_{-0.05}$& $0.72^{+0.42}_{-0.33}$&$2.6^{+0.5}_{-0.9}$&43/50\\
Mrk 273 & $2500^{+1500}_{-890}$&$3.2^{+2.9}_{-1.4}$&$6.5^{+0.21}_{-0.17}$&0.05f&$970^{+630}_{-680}$&$4.8\pm0.91$&&1.0f&$0.35^{+0.082}_{-0.097}$&$0.84^{+0.48}_{-0.24}$&$0.28^{+.17}_{-0.097}$&$1.3^{+0.15}_{-0.19}$&90/68\\
Mrk 463 &$ 2800^{+1300}_{-850}$&$ 5.8^{+5.0}_{-2.7}$&\nodata&\nodata&$<960$&$ 6.6\pm1.3$&&2.1f&$ 0.40^{+0.085}_{-0.094}$&$ 0.66^{+0.35}_{-0.39}$&$ 0.16^{+0.086}_{-0.084}$&$ 1.1\pm0.17$& 96/87  \\
NGC 7582 (1994) &  $1000\pm70$& $84^{+6.4}_{-5.9}$& $6.2^{+0.18}_{-0.36}$ & 0.05f &$160^{+100}_{-84}$&$140^{+6.9}_{-6.1}$&&$5.3^{+5.7}_{-1.6}$&$1.5^{+0.44}_{-0.32}$&$0.46^{+0.32}_{-0.19}$ & $0.56^{+0.61}_{-0.32}$&$4.10^{+0.9}_{-1.0}$&365/297\\ 
NGC 7582 (1996)& $1400\pm50$& $93^{+4.6}_{-4.2}$& $6.3\pm0.08$ & 0.05f &$140^{+55}_{-48}$&$140^{+4.3}_{-4.0} $&&$5.6^{+3.4}_{-1.4} $&$1.9\pm0.20$&$0.60^{+0.24}_{-0.30}$&$0.32^{+0.16}_{-0.17}$ &$4.2\pm0.4$&538/480 \\ 

\enddata

\tablenotetext{a}{Column density in units of $10^{20}{\rm\,cm^{-2}}$.}
\tablenotetext{b}{Normalization of power law in units of
$10^{-4} {\rm\,photons\, keV^{-1}\,cm^{-2}\,s^{-1}}$ at 1 keV.}
\tablenotetext{c}{Energy of line center in keV.}
\tablenotetext{d}{Line width in keV.}
\tablenotetext{e}{Equivalent width of line in eV.}
\tablenotetext{f}{2.0--10.0 keV model flux in  SIS0 detector in units of
$10^{-13}{\rm\,erg\,cm^{-2}\,s^{-1}}$.}
\tablenotetext{g}{Normalization of the soft power-law component in units of 
$10^{-4} {\rm\,photons\, keV^{-1}\,cm^{-2}\,s^{-1}}$ at 1 keV.}
\tablenotetext{h}{Temperature of thermal plasma in keV.}
\tablenotetext{i}{Normalization of thermal component in units of $10^{-4}\times K$, 
where $K=(10^{-14}/(4\pi D^2))\int n_e n_H dV, D$ is the distance to the source (cm), 
$n_e$ is the electron density (${\rm cm^{-3}}$), and $n_H$ is the hydrogen density 
(${\rm cm^{-2}}$).}
\tablenotetext{j}{0.5--2.0 keV model flux in SIS0 detector in units of
$10^{-13}{\rm\,erg\,cm^{-2}\,s^{-1}}$.}
\tablecomments{Power law photon index is fixed at 1.9.
Additional fixed parameters are marked with f.   
Errors are 90\% confidence limits for two interesting parameters, except
fluxes, where errors are 90\% confidence for one parameter.
Parameters that are constrained by hard limits are marked with a colon.}

\end{deluxetable}

\hoffset 0in
\thispagestyle{empty}
\begin{deluxetable}{lrccrccccccc}
\tabletypesize{\small}
\tablewidth{0pt}
\tablecaption{X-ray Luminosity\label{tab:xlum}}
\tablehead{
&\multicolumn{2}{c}{HRI}&&\multicolumn{2}{c}{PSPC}&&\multicolumn{5}{c}{ASCA} \\
\cline{2-3} \cline{5-6} \cline{8-12} \\
\colhead{Galaxy}
&\colhead{Extended $SX$\tablenotemark{a}}&\colhead{Total $SX$\tablenotemark{b}}
&&\colhead{Extended $SX$\tablenotemark{a}}&\colhead{Total $SX$\tablenotemark{b}}
&&\colhead{Thermal $SX$\tablenotemark{c}}&\colhead{Total $SX$\tablenotemark{b}}
&\colhead{Thermal $MX$\tablenotemark{d}}&\colhead{Total $MX$\tablenotemark{e}}
&\colhead{$HX$\tablenotemark{f}}
}
\startdata
NGC 1068 &            41.4 &     41.7 &&     41.3 &     41.7  &&   41.4 & 41.5  &  41.4  &    41.6 &  41.1  \\
Mrk 1066 &         $<$39.8 &     40.6 && \nodata  & \nodata   &&   40.5 & 40.7  &  40.4  &    40.9 &  41.0  \\
Mrk 1073 &            40.7 &     41.0 && \nodata  & \nodata   &&   41.1 & 41.1  &  41.1  &    41.5 &  41.7  \\
Mrk 78   &            41.1 &     41.5 &&  $<$41.2 &     41.4  &&\nodata &\nodata&\nodata & \nodata &\nodata \\
IC 3639  &            40.1 &     40.5 && \nodata  & \nodata   &&   40.5 & 40.5  & 40.7   &    40.8 &  41.0  \\
NGC 5135 &            40.8 &     41.2 &&     40.5 &     41.2  &&   40.8 & 41.1  & 40.8   &    41.2 &  41.4  \\
Mrk 266  &            41.6 &     41.7 &&     41.4 &     41.7  &&   41.3 & 41.6  & 41.3   &    41.8 &  41.5  \\
Mrk 273  &         $<$41.0 &     41.4 &&  $<$40.8 &     41.4  &&   41.2 & 41.6  & 41.2   &    41.8 &  42.1  \\
Mrk 463  &        \nodata  & \nodata  &&  $<$41.5 &     41.9  &&   41.2 & 41.7  & 41.2   &    42.0 &  42.5  \\
Mrk 477  &         $<$41.2 &     41.8 && \nodata  & \nodata   &&\nodata & 41.5  &\nodata &    41.9 &  42.5  \\
NGC 6221 &            40.5 &     40.8 && \nodata  & \nodata   &&   39.9 & 41.1  & 39.9   &    41.6 &  41.8  \\
NGC 7130 &            40.8 &     41.1 && \nodata  & \nodata   &&   40.8 & 41.0  & 40.8   &    41.2 &  41.0  \\
NGC 7582 &            40.1 &     40.3 &&     39.9 &     40.3  &&   39.9 & 40.4  & 39.6   &    41.2 &  41.9  \\
NGC 7674 &        \nodata  & \nodata  &&     41.3 &     41.5  &&\nodata &\nodata&\nodata & \nodata &\nodata \\
\enddata                    
\tablenotetext{a}{Extended soft X-ray luminosity (0.5-2.0 keV), 
corrected for Galactic absorption.}
\tablenotetext{b}{Total soft X-ray luminosity (0.5-2.0 keV), 
corrected for Galactic absorption.}
\tablenotetext{c}{Thermal soft X-ray luminosity (0.5-2.0 keV), 
corrected for Galactic absorption.}
\tablenotetext{d}{Thermal medium X-ray luminosity (0.5-4.5 keV), 
corrected for Galactic absorption.}
\tablenotetext{e}{Total medium X-ray luminosity (0.5-4.5 keV), 
corrected for Galactic absorption.}
\tablenotetext{f}{Total observed hard X-ray luminosity (2.0-10 keV).}
\tablecomments{All quantities are log erg s$^{-1}$.}
\end{deluxetable}

\begin{deluxetable}{lccccc}
\tablewidth{0pt}
\tablecaption{Metallicity Effects\label{tab:lowz}}
\tablehead{
\colhead{Galaxy}&$f_{scatt}$&$f_{scatt}$&\colhead{$f$\tablenotemark{a}$_{therm}$}
&\colhead{$f$\tablenotemark{a}$_{therm}$}&\colhead{$\chi^2/$dof}\\
&\colhead{$Z=Z_\odot$}&\colhead{$Z=0.05Z_\odot$}
&\colhead{$Z=Z_\odot$}&\colhead{$Z=0.05Z_\odot$}
&\colhead{$Z=0.05Z_\odot$}
}
\startdata
NGC 1068       &\nodata &\nodata & 0.80& 0.81  &2902/908\\
Mrk 1066       &\nodata &\nodata & 0.51& 0.86  & 75/73  \\
Mrk 1073       &\nodata &\nodata & 0.91& 0.98  & 118/63 \\
IC 3639        &\nodata &\nodata & 1.00& 1.00  & 52/53  \\
NGC 5135       & 0.02   & 0.03   & 0.54& 0.76  & 135/119\\
Mrk 266        &\nodata &\nodata & 0.43& 0.94  & 29/49  \\
Mrk 273        & 0.11   &\nodata & 0.43& 1.00  & 94/69  \\
Mrk 463        & 0.07   & 0.10   & 0.31& 0.35  & 102/90 \\
Mrk 477        & 0.07   &\nodata & 0.00&\nodata&\nodata \\
NGC 6221       & 0.07   & 0.01   & 0.05& 0.22  & 798/765\\
NGC 7130       &\nodata &\nodata & 0.57& 0.57  & 75/74  \\
NGC 7582 (1994)& 0.02   & 0.02   & 0.16& 0.27  & 362/296\\
NGC 7582 (1996)& 0.02   & 0.02   & 0.28& 0.22  & 541/479\\
\enddata                    
\tablenotetext{a}{Thermal fraction of soft emission.} 
\end{deluxetable}

\centering

\end{document}